\documentclass[]{emulateapj}
\usepackage{natbib}
\usepackage{apjfonts}
\usepackage{epsfig}

\def\mag      {{\ifmmode~\mathrm{mag}\else$~\mathrm{mag}$\fi} } 

\def\lb{\left[}
\def\rb{\right]}

\newcommand{\ave}[1]{\langle #1 \rangle}

\newcommand{\Mpc}{\mathrm{Mpc}}
\newcommand{\Msun}{M_{\sun}}
\newcommand{\mstar}{m_{\star}}
\newcommand{\Omegam}{\Omega_{\mathrm{m}}}
\newcommand{\Omegab}{\Omega_{\mathrm{b}}}

\newcommand{\Omegal}{\Omega_{\Lambda}}
\newcommand{\rhoc}{\rho_{\mathrm{c}}}

\newcommand{\Xp}{X_{\mathrm{p}}}
\newcommand{\Yp}{Y_{\mathrm{p}}}
\newcommand{\alphaB}{\alpha_{\mathrm{B}}}

\newcommand{\QHII}{Q_{\mathrm{HII}}}
\newcommand{\CHII}{C_{\mathrm{HII}}}
\newcommand{\dQHIIdt}{\dot{Q}_{\mathrm{HII}}}
\newcommand{\nH}{n_{\mathrm{H}}}
\newcommand{\trec}{t_{\mathrm{rec}}}
\newcommand{\dniondt}{\dot{n}_{\mathrm{ion}}}
\newcommand{\xiion}{\xi_{\mathrm{ion}}}
\newcommand{\etasfr}{\eta_{\mathrm{sfr}}}
\newcommand{\fesc}{f_{\mathrm{esc}}}
\newcommand{\fmax}{f_{\mathrm{max}}}
\newcommand{\sigmaT}{\sigma_{\mathrm{T}}}

\newcommand{\rhostar}{\rho_{\star}}
\newcommand{\rhoUV}{\rho_{\mathrm{UV}}}
\newcommand{\rhoUVA}{\rho_{\mathrm{UV},z=4}}
\newcommand{\rhoUVB}{\rho_{\mathrm{UV},z=7}}
\newcommand{\rhoSFR}{\rho_{\mathrm{SFR}}}
\newcommand{\MUV}{M_{\mathrm{UV}}}
\newcommand{\AV}{A_{\mathrm{V}}}
\newcommand{\tauV}{\tau_{\mathrm{V}}}
\newcommand{\YWFC}{Y_{\mathrm{105}}}
\newcommand{\JAWFC}{J_{\mathrm{125}}}
\newcommand{\JBWFC}{J_{\mathrm{140}}}
\newcommand{\HWFC}{H_{\mathrm{160}}}
\newcommand{\phistar}{\phi_{\star}}
\newcommand{\Mstar}{M_{\star}}

\newcommand{\eV}{\mathrm{eV}}
\newcommand{\xiionunit}{\mathrm{ergs}^{-1}~\mathrm{Hz}}
\newcommand{\rhoUVunit}{\mathrm{ergs}~\mathrm{s}^{-1}~\mathrm{Hz}^{-1}~\mathrm{Mpc}^{-3}}
\newcommand{\LUV}{L_{\mathrm{UV}}}
\newcommand{\LUVunit}{\mathrm{ergs}~\mathrm{s}^{-1}~\mathrm{Hz}^{-1}}
\newcommand{\etaSFRunit}{M_{\sun}~\mathrm{yr}^{-1}~\mathrm{ergs}^{-1}~\mathrm{s}~\mathrm{Hz}}

\shorttitle{UDF12: New Constraints on Cosmic Reionization}
\shortauthors{Robertson et al.}

\begin{document}

\title{New Constraints on Cosmic Reionization from the 2012 Hubble Ultra Deep Field Campaign}

\author{Brant E. Robertson\altaffilmark{1},
Steven R. Furlanetto\altaffilmark{2},
Evan Schneider\altaffilmark{1},
Stephane Charlot\altaffilmark{3},
Richard S. Ellis\altaffilmark{4},
Daniel P. Stark\altaffilmark{1},
Ross J. McLure\altaffilmark{5},
James S. Dunlop\altaffilmark{5},
Anton Koekemoer\altaffilmark{6},
Matthew A. Schenker\altaffilmark{4},
Masami Ouchi\altaffilmark{7},
Yoshiaki Ono\altaffilmark{7},
Emma Curtis-Lake\altaffilmark{5},
Alexander B. Rogers\altaffilmark{5},
Rebecca A. A. Bowler\altaffilmark{5},
Michele Cirasuolo\altaffilmark{5}
}

\altaffiltext{1}{Department of Astronomy and Steward Observatory, University of Arizona, Tucson AZ 85721}
\altaffiltext{2}{Department of Physics \& Astronomy, University of California, Los Angeles CA 90095}
\altaffiltext{3}{UPMC-CNRS, UMR7095, Institut d'Astrophysique de Paris, F-75014, Paris, France}
\altaffiltext{4}{Department of Astrophysics, California Institute of Technology, MC 249-17, Pasadena, CA 91125} 
\altaffiltext{5}{Institute for Astronomy, University of Edinburgh, Royal Observatory, Edinburgh EH9 3HJ, UK}
\altaffiltext{6}{Space Telescope Science Institute, Baltimore, MD 21218}
\altaffiltext{7}{Institute for Cosmic Ray Research, University of Tokyo, Kashiwa City, Chiba 277-8582, Japan} 

\begin{abstract}
Understanding cosmic reionization requires the identification and characterization of early sources of
hydrogen-ionizing photons. The 2012 Hubble Ultra Deep Field (UDF12) campaign has acquired the deepest infrared images 
with the Wide Field Camera 3 aboard {\it Hubble Space Telescope} and, for the first time, systematically
explored the galaxy population deep into the era when cosmic microwave background (CMB) data indicates 
reionization was underway. The UDF12 campaign  thus provides the best constraints to date on the abundance, 
luminosity distribution, and spectral properties of early star-forming galaxies. We synthesize the new UDF12 results 
with the most recent constraints from CMB observations to infer redshift-dependent ultraviolet (UV) luminosity densities, 
reionization histories, and electron scattering optical depth evolution consistent with the available data.  
Under reasonable assumptions about the escape
fraction of hydrogen ionizing photons and the intergalactic medium
clumping factor, we find that 
to fully reionize the universe by redshift $z\sim6$ 
the population of star-forming galaxies at
redshifts $z\sim7-9$ likely must extend in luminosity below the UDF12 limits to absolute UV magnitudes of $\MUV\sim-13$ or 
fainter. 
Moreover, low levels of star formation extending to redshifts $z\sim15-25$, as suggested by the 
normal UV colors of $z\simeq$7-8 galaxies and the smooth decline in abundance with redshift 
observed by UDF12 to $z\simeq10$, are additionally likely required to reproduce the optical depth to electron scattering inferred 
from CMB observations.  
\end{abstract}

\keywords{cosmology:reionization -- galaxies:evolution -- galaxies:formation}

%
%
\section{Introduction}
\label{section:intro}

The process of cosmic reionization remains one of the most important outstanding problems
in galaxy formation and cosmology. After recombination at $z\approx1090$ \citep{hinshaw2012a}, gas
in the universe was mostly neutral. However, 
observations of the 
\citet{gunn1965a} trough in quasar spectra 
\citep[e.g.,][]{fan2001a,fan2002a,fan2003a,fan2006a,djorgovski2001a} indicate that 
intergalactic gas has become almost fully reionized by redshift $z\sim5$.  The electron
scattering optical depth inferred from CMB observations suggests that if the universe was
instantaneously reionized, then reionization would occur as early as redshift $z\approx10$
\citep{spergel2003a,hinshaw2012a}.  Given the dramatic decline in the abundance of
quasars beyond redshift $z\sim6$, they very likely cannot be a significant contributor
to cosmic reionization \citep[e.g.,][]{willott2010a,fontanot2012a} even though quasars have
been discovered as early as $z\sim7$ \citep{mortlock2011a}.
Star-forming galaxies at redshifts $z\gtrsim6$ have therefore long been postulated as the
likely agents of cosmic reionization, and their time-dependent abundance and
spectral properties are thus crucial ingredients for understanding how intergalactic hydrogen
became reionized \citep[for reviews, see][]{fan2006b,robertson2010a,loeb-furlanetto-2012}.

We present an analysis of the implications of the 2012 Hubble Ultra Deep Field\footnote{http://udf12.arizona.edu} 
(UDF12) campaign results on the abundance and
spectral characteristics of galaxies at $z\sim7-12$ for the reionization process.
The UDF12 campaign is a 128-orbit {\it Hubble Space Telescope} ({\it HST}) program (GO 12498,
PI: Ellis) with the infrared (IR) channel on the Wide Field Camera 3 (WFC3/IR) that
acquired the deepest ever images in the IR with {\it HST} in the Hubble Ultra Deep
Field (HUDF) in Fall 2012 
(the UDF12 project and data overview are described in \citealt{ellis2012a} and
\citealt{koekemoer2012a}).
Combined with previous HUDF observations (GO 11563, PI: G. Illingworth; 
GO 12060, 12061, 12062, PIs: S. Faber and H. Ferguson; GO 12099, PI: A. Riess), the
UDF12 imaging reaches depths of $\YWFC=30$, $\JAWFC=29.5$, $\JBWFC=29.5$, and $\HWFC=29.5$
($5-\sigma$ AB magnitudes).  The UDF12 observations have provided the first determinations
of the galaxy abundance at redshifts $8.5\le z\le 12$ \citep{ellis2012a}, precise determinations
of the galaxy luminosity function at redshifts $z\sim7-8$ \citep{schenker2012a,mclure2012a},
robust ultraviolet (UV) spectral slope measurements at $z\sim7-8$ \citep{dunlop2012b}, and 
size-luminosity relation measures at redshifts $z\sim6-8$ \citep{ono2012a}.

Our earlier UDF12 publications already provide some new constraints on the
role that galaxies play in cosmic reionization and the duration of the process.
In \citet{ellis2012a}, we argued that continuity in the declining abundance of
star-forming galaxies over $6<z<10$ (and possibly to $z\simeq12$), implied
the likelihood of further star formation beyond the redshift limits currently
probed. Likewise, in \citet{dunlop2012b}, the constancy of the UV continuum
slope measured in $z\simeq7-9$ galaxies over a wide range in luminosity
supports the contention that the bulk of the stars at this epoch are already
enriched by earlier generations. Collectively, these two results support
an extended reionization process.

We synthesize these UDF12 findings with the recent 9-year {\it Wilkinson Microwave Anisotropy
Probe} ({\it WMAP}) results \citep{hinshaw2012a} and stellar mass density measurements \citep{stark2012a}
to provide new constraints on the role of high-redshift star-forming
galaxies in the reionization process.  Enabled by the new observational findings, we
perform Bayesian inference using a simple parameterized model for the evolving UV luminosity
density to find reionization histories, stellar mass density evolutions, and
electron scatter optical depth progressions consistent
with the available data.  We limit the purview of this paper to empirical modeling of the
reionization process, and comparisons with more detailed galaxy formation models will
be presented in a companion paper (Dayal et al., in preparation).

Throughout this paper, we assume the 9-year {\it WMAP} cosmological parameters 
(as additionally constrained by external CMB datasets;
$h=0.705$, $\Omegam=0.272$, $\Omegal=0.728$, $\Omegab=0.04885$).  Magnitudes are reported
using the AB system \citep{oke1983a}.  All Bayesian inference and maximum likelihood fitting
is performed using the {\it MultiNest} code \citep{feroz2008a,feroz2009a}.

%
%
\section{The Process of Cosmic Reionization}
\label{section:reionization}

Theoretical models of the reionization process have a long history.
Early analytic and numerical
models of the reionization process 
\citep[e.g.,][]{madau1999a,miralda-escude2000a,gnedin2000a,barkana2001a,razoumov2002a,wyithe2003a,ciardi2003a}
highlighted the essential physics that give rise to the ionized intergalactic
medium (IGM) at late times.  In the following description of the cosmic reionization process
we follow most closely the modeling of \citet{madau1999a}, \citet{bolton2007a}, \citet{robertson2010a}, and \citet{kuhlen2012a}, but there has been closely related recent work 
by \citet{ciardi2012a} and \citet{jensen2013a}.

The reionization process is a balance between
the recombination of free electrons with protons to form neutral hydrogen and
the ionization of hydrogen atoms by cosmic Lyman continuum photons with energies
$E > 13.6~\eV$.  The dimensionless volume filling fraction of ionized hydrogen $\QHII$ can be expressed
as a time-dependent differential equation capturing these competing effects as
\begin{equation} 
\label{eqn:QHII}
\dQHIIdt = \frac{\dniondt}{\ave{\nH}} - \frac{\QHII}{\trec}
\end{equation} 
\noindent
where dotted quantities are time derivatives.

The comoving density of hydrogen atoms
\begin{equation}
\ave{\nH} = \Xp \Omegab \rhoc
\end{equation}
\noindent
depends on the primordial mass-fraction of hydrogen $\Xp=0.75$ \citep[e.g.,][]{hou2011a},
the critical density $\rhoc = 1.8787 \times 10^{-29} h^{-2}~\mathrm{g}~\mathrm{cm}^{3}$,
and the fractional baryon density $\Omegab$.

As a function of redshift, the average recombination time in the IGM is
\begin{equation}
\label{eqn:trec}
\trec = \lb \CHII \alphaB(T) (1 + \Yp/4\Xp) \ave{\nH} (1+z)^{3} \rb^{-1},
\end{equation}
\noindent
where $\alphaB(T)$ is the case B recombination coefficient for hydrogen (we assume
an IGM temperature of $T=20,000$K), $\Yp=1-\Xp$ is the primordial helium abundance
(and accounts for the number of free electrons per proton in the fully ionized
IGM, e.g., \citealt{kuhlen2012a}), and $\CHII \equiv \ave{\nH^{2}}/\ave{\nH}^{2}$
is the ``clumping factor'' that accounts for the effects of IGM inhomogeneity through
the quadratic dependence of the
recombination rate on density. 

Simulations suggest that the clumping factor 
of IGM gas is $\CHII\approx1-6$ at the redshifts of interest 
\citep[e.g.,][]{sokasian2003a,iliev2006a,pawlik2009a,shull2012a,finlator2012a}.
While early
hydrodynamical simulation studies suggested that the clumping factor could
be as high as $\CHII\sim10-40$ at redshifts $z<8$ \citep[e.g.,][]{gnedin1997a}, 
recent studies that 
separately identify IGM and interstellar medium gas in the simulations and
employ a more detailed modeling of the evolving UV background have found lower
values of $\CHII$.  An interesting study of the redshift evolution of the
clumping factor was provided by \citet[][see their Figures 5 and 7]{pawlik2009a}. 
At early times in their simulations ($z\geq10$), the clumping factor was low ($\CHII<3$) but increased
with decreasing redshift at a rate similar to predictions from the \citet{miralda00}
model of the evolving IGM density probability distribution function.  In the absence
of photoheating, at lower redshifts ($z\lesssim9$) the clumping factor would begin 
to increase more rapidly than the \citet{miralda00} prediction to reach $\CHII\sim10-20$
by $z\sim6$.  In the presence of photoheating, the evolution of the clumping factor
depends on the epoch when the uniform UV background becomes established. If the
IGM was reheated early ($z\sim10-20$) the predicted rise in clumping factor breaks after
the UV background is established and increases only slowly to $\CHII\sim3-6$ at 
late times ($z\sim6$).
If instead, and perhaps more likely, the IGM is reheated later (e.g, $z\sim6-8$) the
clumping factor may actually decrease at late times from $\CHII\sim6-10$ at $z\sim8$
to $\CHII\sim3-6$ at $z\sim6$.  

The results of these simulations \citep[e.g.,][]{pawlik2009a,finlator2012a} in part motivate
our choice to treat the clumping factor as a constant $\CHII\sim3$ since over a wide range 
of possible redshifts for the establishment of the UV background the clumping factor is
expected to be $\CHII\sim2-4$ at $z\lesssim12$ \citep[see Figure 5 of ][]{pawlik2009a} 
and lower at earlier times.  
In comparison with our previous work 
\citep{robertson2010a}, where we considered $\CHII=2-6$ and frequently used 
$\CHII=2$ in Equation \ref{eqn:trec}, we will see that our models complete reionization
somewhat later when a somewhat larger value of $\CHII$ is more appropriate.
However, we note that the end of the reionization process may be more complicated than
what we have described above
(see, e.g., Section 9.2.1 of \citealt{loeb-furlanetto-2012}). As reionization progresses, 
the ionized phase penetrates more and more deeply into dense clumps within the IGM -- 
the material that will later form the Lyman-$\alpha$ forest (and higher column density systems).  
These high-density clumps recombine much faster than average, so $\CHII$ may increase throughout 
reionization \citep{furl05-rec}.  Combined with the failure of Equation \ref{eqn:QHII} to model
the detailed distribution of gas densities in the IGM, we expect our admittedly crude approach 
to fail at the tail end of reionization. Fortunately, we are primarily 
concerned with the middle phases of reionization here, so any unphysical behavior when 
$\QHII$ is large is not important for us.  

The comoving production rate $\dniondt$
of hydrogen-ionizing photons available to reionize the IGM depends on the intrinsic productivity
of Lyman continuum radiation by stellar populations within galaxies parameterized in terms of the
rate of hydrogen-ionizing photons per unit UV (1500$\mathring{A}$) 
luminosity $\xiion$ (with units of $\xiionunit$), 
the fraction $\fesc$ of such photons that escape to affect the IGM, and the total UV luminosity
density $\rhoUV$ (with units of $\rhoUVunit$) supplied by star-forming galaxies to some limiting
absolute UV magnitude $\MUV$.  The product 
\begin{equation}
\label{eqn:dniondt}
\dniondt = \fesc \xiion \rhoUV
\end{equation}
then determines the newly available number density of Lyman continuum photons per second capable
of reionizing intergalactic hydrogen. 
We note that the expression of $\dniondt$ in terms of UV luminosity density
rather than star formation rate \citep[c.f.,][]{robertson2010a} is largely a matter of choice; 
stellar population synthesis models with assumed star formation histories are required to estimate
$\xiion$ and using the star formation rate density $\rhoSFR$ in Equation \ref{eqn:dniondt} therefore
requires no additional assumptions.  Throughout this paper, we choose $\fesc=0.2$.  As shown by
\citet{ouchi2009a}, escape fractions comparable 
to or larger than $\fesc=0.2$ during the reionization
epoch are required for galaxies with typical stellar populations 
to contribute significantly.  We also consider an evolving
$\fesc$ with redshift, with the results discussed in Section \ref{section:ionizing_background} below. 

The advances presented in this paper come primarily from the new UDF12
constraints on the abundance of star-forming galaxies over $6.5<z<12$, the
luminosity functions down to $\MUV\simeq-17$, and robust determinations
of their UV continuum colors. For the latter, in Section \ref{section:beta}, we use the 
UV spectral slope of high-redshift galaxies by \citet{dunlop2012b}
and the stellar population synthesis models of \citet{bruzual2003a}
to inform a choice for the number $\xiion$ of ionizing photons produced per unit
luminosity. For the former, the abundance and luminosity distribution of high-redshift galaxies 
determined by \citet{ellis2012a}, \citet{schenker2012a}, and \citet{mclure2012a} provide 
estimates of the evolving UV luminosity density $\rhoUV$.
The evolving UV luminosity density supplied by star-forming galaxies brighter than
some limiting magnitude $\MUV$ is simply related to
an integral of the luminosity function as
\begin{equation}
\label{eqn:rhoUV}
\rhoUV(z) = \int_{-\infty}^{\MUV} \Phi(M) L(M) dM,
\end{equation}
\noindent
where $L$ is the luminosity and the functional form of the galaxy luminosity function is often assumed
to be a \citet{schechter1976a} function
\begin{equation}
\Phi(M) = 0.4 \ln 10~\phistar \left[10^{0.4(\Mstar -M)}\right]^{1+\alpha} \exp\left[-10^{0.4(\Mstar -M)}\right]
\end{equation}
parameterized in terms of the normalization $\phistar$ (in units of $\Mpc^{-3}\mag^{-1}$), the characteristic galaxy
magnitude $\Mstar$, and the faint-end slope $\alpha$.  Each of these parameters may evolve with redshift $z$, which
can affect the relative importance of faint galaxies for reionization \citep[e.g.,][]{oesch2009a,bouwens2012a}.  
In Sections \ref{section:rho_uv} and \ref{section:rho_uv_like} 
below, we present our method
of using the previous and UDF12 data sets to infer 
constraints on the luminosity density as a function of redshift and limiting magnitude.

\subsection{Stellar Mass Density as a Constraint on Reionization}
\label{section:rhostar}

The UV luminosity density is supplied by short-lived, massive stars and therefore
reflects the time rate of change of the stellar mass density $\rhostar(z)$.
In the context of our model, there are two routes for estimating the stellar mass
density.  First, we can integrate the stellar mass density supplied by the star formation
rate inferred from the evolving UV luminosity density as
\begin{equation}
\label{eqn:rhostar_int}
\rhostar(z) =  (1-R) \int_{\infty}^{z} \etasfr(z') \rhoUV(z') \frac{dt}{dz'} dz',
\end{equation}
\noindent
where $\etasfr(z)$ provides the stellar population model-dependent conversion between UV 
luminosity and star formation rate (in units $\etaSFRunit$), $R$ is the fraction of
mass returned from a stellar population to the ISM ($28\%$ for a \citealt{salpeter1955a}
model after $\sim10$~Gyr for a $0.1-100\Msun$ IMF), and $dt/dz$ gives the rate of
change of universal time per unit redshift (in units of $\mathrm{yr}$).  While $\etasfr(z)$
is in principle time-dependent, there is no firm evidence yet of its evolution and we adopt a
constant value throughout.

While the evolving stellar mass density can be calculated in our model, the observational
constraints on $\rhostar$ have involved integrating a composite stellar mass function
constructed from the UV luminosity function and a stellar mass to UV luminosity relation
\citep[][see also \citealt{labbe2012a}]{gonzalez2011a,stark2012a}. Using near-IR observations with 
the
{\it Spitzer Space Telescope}, stellar masses of UV-selected galaxies are measured as
a function of luminosity.  \citet{stark2012a} find that the stellar mass -- $\LUV$ relation, 
corrected for nebular emission contamination, is well described by
\begin{equation}
\label{eqn:Mstar_MUV}
\log \mstar = 1.433 \log \LUV -31.99 + f_{\star}(z)
\end{equation}
\noindent
where $\mstar$ is the stellar mass in $\Msun$, $\LUV$ is the UV luminosity density in
$\LUVunit$, and $f_{\star}(z) = [0, -0.03, -0.18, -0.40]$ for redshift 
$z \approx [4, 5, 6, 7]$.

The stellar mass density will then involve an integral over the product of the
UV luminosity function and the stellar mass $\mstar(\LUV)$.  However, the significant
scatter in the $\mstar(\LUV)$ relation 
\citep[$\sigma\approx0.5$ in $\log \mstar$, see][]{gonzalez2011a,stark2012a}
must be taken into account.  The gaussian scatter $p[M'-M(\log \mstar)]$ in luminosity contributing
at a given $\log \mstar$ can be incorporated into the stellar mass function $dn_{\star}/d\log\mstar$
with a convolution over the UV luminosity function. 
We write the stellar mass function as
\begin{equation}
\label{eqn:mstar_function}
\frac{dn_{\star}}{d\log\mstar} = \frac{d M}{d\log \mstar} \int_{-\infty}^{\infty} \Phi(M') p[M'-M(\log \mstar)] d M'.
\end{equation}
\noindent
For vanishing scatter Equation \ref{eqn:mstar_function} would give simply
$dn_{\star}/d\mstar = \Phi[M(\mstar)] \times dM/d\mstar$.
The stellar mass density $\rhostar$ can be computed by integrating this mass function as
\begin{equation}
\label{eqn:rhostar_conv}
\rhostar(<\mstar, z) = \int_{-\infty}^{\log \mstar} \frac{dn_{\star}}{d\log\mstar'} \mstar' d\log \mstar'.
\end{equation}
\noindent
A primary feature of the stellar mass function is that the stellar mass -- UV luminosity relation
and scatter
flattens it relative to the UV luminosity function.  Correspondingly, the stellar mass density
converges faster with decreasing stellar mass or luminosity than does the UV luminosity density 
(Equation \ref{eqn:rhoUV}).  The stellar mass density then serves as an additional, 
integral constraint on $\dniondt$. 

\subsection{Electron Scattering Optical Depth}
\label{section:tau}

Once the evolving production rate of ionizing photons $\dniondt$ is determined, 
the reionization history $\QHII(z)$ of the universe can be calculated by integrating 
Equation \ref{eqn:QHII}.  An important integral constraint on the reionization history
is the electron scattering optical depth $\tau$ inferred from observations of the CMB.
The optical depth can be calculated from the reionization history as a function of 
redshift $z$ as
\begin{equation}
\label{eqn:tau}
\tau(z) = \int_{0}^{z} c \ave{\nH} \sigmaT f_{\mathrm{e}} \QHII(z') H(z') (1+z')^{2} dz',
\end{equation}
where $c$ is the speed of light, $\sigmaT$ is the Thomson cross section, and $H(z)$ is the
redshift-dependent Hubble parameter.  The number $f_{\mathrm{e}}$ of free electrons per 
hydrogen nucleus in the ionized IGM depends on the ionization state of helium.  Following
\citet{kuhlen2012a} and other earlier works, we assume that helium is doubly ionized ($f_{\mathrm{e}} = 1+\Yp/2\Xp$)
at $z\le4$ and singly ionized ($f_{\mathrm{e}} = 1+\Yp/4\Xp$) at higher redshifts. To utilize
the observational constraints on $\tau$ as a constraint on the reionization history, we employ
the posterior probability distribution $p(\tau)$ determined from the Monte Carlo Markov Chains
used in the 9-year {\it WMAP} results\footnote{http://lambda.gsfc.nasa.gov} as a 
marginalized likelihood for our derived $\tau$ values.
This method is described in more detail in Section \ref{section:constraints}.

%
%
\section{UV Spectral Slopes and the Ionizing Photon Budget}
\label{section:beta}

A critical ingredient for determining the comoving production rate of 
hydrogen ionizing photons is the ratio $\xiion$ of the Lyman continuum 
photon emission rate per
unit UV (1500$\mathring{A}$) luminosity spectral density of individual sources. Since
Lyman continuum photons are predominately produced by hot, massive, 
UV-bright stars, it is sensible to expect that $\xiion$ will be
connected with the UV spectral slope of a stellar population.  Here,
we use observational constraints on the UV slope of high-redshift galaxies
determined from the UDF12 campaign by \citet{dunlop2012b} and stellar
population synthesis models by \citet[][BC03]{bruzual2003a} to estimate
a physically-motivated $\xiion$ consistent with the data.
In this paper, we concentrate on placing constraints on $\xiion$; for a much more 
detailed analysis and interpretation of the UV slope results from UDF12, please see \citet{dunlop2012b}.

\begin{figure*}
\figurenum{1}
\epsscale{1.1}
\plotone{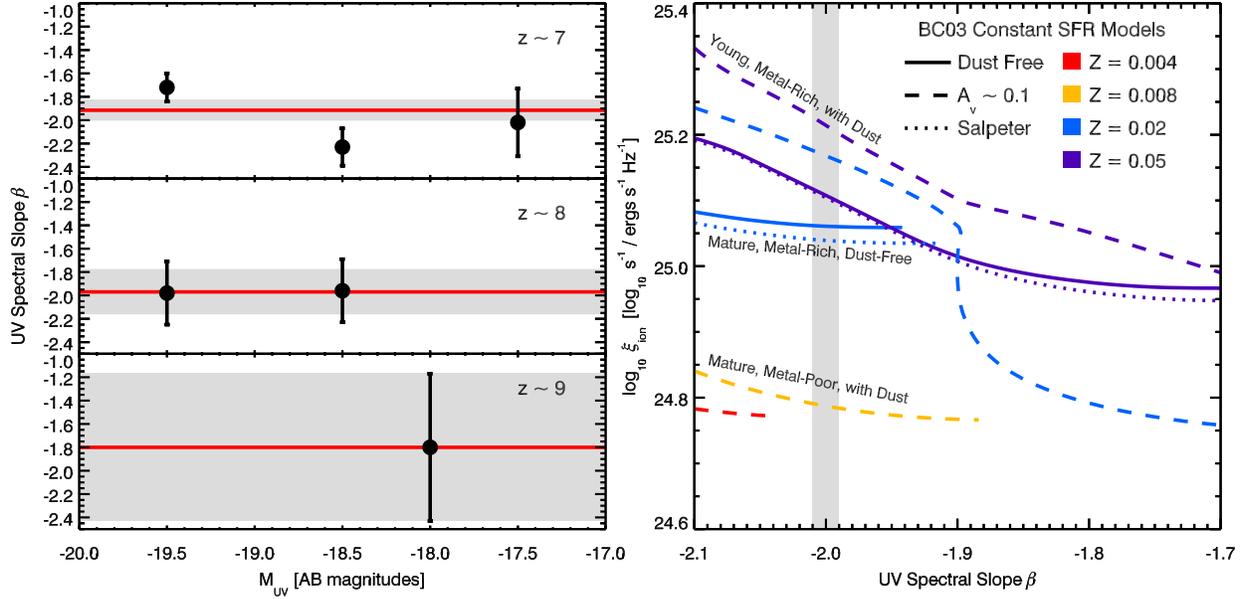}
\caption{\label{fig:beta}
Spectral properties of high-redshift galaxies and the corresponding properties of stellar populations.
\citet{dunlop2012b} used the new UDF12 {\it HST} observations to measure the UV spectral slope $\beta$
of $z\sim7-9$ galaxies as a function of luminosity (data points, left panel).  As the data are consistent
with a constant $\beta$ independent of luminosity, we have fit constant values of $\beta$ at redshifts $z\sim 7-8$ 
(maximum likelihood values of $\beta(z\sim7) = -1.915$ and $\beta(z\sim8) = -1.970$ shown as red lines, inner 68\% credibility
intervals shown as grey shaded regions; at $z\sim9$ the line and shaded region reflect the best fit value of 
$\beta(z\sim9) = -1.80\pm0.63$).  The data are broadly consistent with $\beta=-2$ 
(indicated with grey band in right panel), independent of redshift and luminosity.
To translate the UV spectral slope to a ratio $\xiion$ of ionizing photon production rate to UV luminosity, we
use the \citet[][BC03]{bruzual2003a} stellar population synthesis models (right panel) assuming a constant 
star formation rate (SFR).  
The constant SFR models evolve from a declining $\xiion$ with increasing $\beta$ at early times to
a relatively flat $\xiion$ at late times (we plot the values of $\xiion$ vs. $\beta$ for population ages less than the
age of the universe at $z\sim7$, $t=7.8\times10^{8}~\mathrm{yr}$).  
Three broad types of BC03 constant SFR models are 
consistent with values of $\beta=-2$: mature ($\gtrsim10^{8}~\mathrm{yr}$ old), metal-rich, dust free stellar populations,
mature, metal-rich stellar populations with dust ($A_{\mathrm{V}}\sim0.1$ calculated using the \citet{charlot2000a} model), 
and young, metal-rich stellar populations with dust.  Dust free models are plotted with solid lines, while dusty models
are shown as dashed lines.  We assume the \citet{chabrier2003a} initial mass function (IMF), but the \citet{salpeter1955a} IMF
produces similar values of $\xiion$ (dotted lines, dust-free case shown). 
Based on these models we optimistically assume 
$\log \xiion = 25.2~\log~\xiionunit$, but this value is conservative compared with assumptions widely
used in the literature.
}
\end{figure*}

Prior to the UDF12 program, observations of
high-redshift galaxies in the HUDF09 WFC3/IR campaign provided a first
estimate of the UV spectral slopes ($\beta$, where $f_{\lambda}\propto\lambda^{\beta}$) 
of $z\gtrsim7$ galaxies.  Early results
from the HUDF09 team indicated that these high-redshift galaxies 
had extraordinarily blue UV slopes of $\beta\approx-3$ 
\citep{bouwens2010a}, much
bluer than well-studied starburst galaxies at lower redshifts
\citep[e.g.,][]{meurer1999a}.  In the intervening period before the
UDF12 data was acquired, several workers argued against such extreme
values \citep{finkelstein2010a,finkelstein2012a,mclure2011a,dunlop2012a,bouwens2012b,rogers2012a}.
The UDF12 campaign provided significantly deeper $\HWFC$ imaging data used
in the spectral slope determination (e.g., $\beta = 4.43(\JAWFC - \HWFC) -2$)
at redshifts $z\sim7-8$) and added $\JBWFC$ imaging that reduces potential 
observational biases and enables a first UV slope determination at $z\sim9$.
These measurements were presented by \citet{dunlop2012b}, whose results
are discussed in the context of the present paper in Figure \ref{fig:beta} (data points, left panel).  \citet{dunlop2012b}
measured the spectral slope $\beta$ as a function of galaxy luminosity and
redshift in the range $-19.5\le \MUV \le -17.5$ at $z\sim7-8$.  Using their
reported measurements (see their Table 1), we performed simple 
fits of a constant to their $\beta$ values at each redshift separately and found
maximum likelihood values of $\beta(z\sim7) = -1.915$ and $\beta(z\sim8)=-1.970$
(Figure \ref{fig:beta}, red lines in left panel)
consistent with the single $\MUV=-18$ $z\sim9$ measurement of 
$\beta(z\sim9) = -1.80\pm0.63$.   The 68\% credibility intervals on a 
constant $\beta$ at each redshift (Figure \ref{fig:beta}, grey areas in
left panel) suggest that across redshifts $z\sim7-9$ galaxies are consistent
with a non-evolving UV spectral slope in the range $-2.1\le \beta \le -1.7$.
The apparent constancy of $\beta$ with redshift avoids the need for strong 
assumptions about the redshift evolution of galaxy properties.

To connect these UV spectral slope determinations to a value of $\xiion$, we 
must rely on stellar population synthesis models.  We use the standard
BC03 models to
extract model spectral of stellar populations with a range of star formation
histories (bursts and constant star formation rates), metallicities ($Z = 0.0001-0.05$), 
dust absorption ($A_{\mathrm{V}}\approx0.1-1$; calculated using 
the \citealt{charlot2000a} dust model) and 
initial mass functions (IMF; \citealt{chabrier2003a} and \citealt{salpeter1955a}).  
These models provide both 
the Lyman continuum produced by hot stars and the full spectral energy distributions (SEDs)
of the composite stellar population per unit star formation rate (SFR) or stellar mass.
We determine $\xiion$ for each model by dividing the Lyman continuum photon production 
rate per unit SFR or stellar mass by 
a $1500\mathring{A}$ luminosity spectral density (in $\LUVunit$)
per unit SFR or stellar mass
measured with a synthetic filter with a flat response and $100\mathring{A}$ width.  The
model SFR or stellar mass then scales out of the 
ratio $\xiion$ (with units $\xiionunit$).
By measuring the UV slope values of each model using synthetic photometry with the $\JAWFC$ and
$\HWFC$ total throughput response on SEDs redshifted appropriately for $z\sim7$ observations, 
the value of $\xiion$ for each model $\beta$ can be studied
as a function of age, metallicity, IMF, and star formation history.  We have checked that
our methods for measuring $\beta$ and IR colors from the synthetic model spectra reproduce results from
the literature (see \citealt{robertson2007a} and Figure 2 of \citealt{rogers2012a}).

The right panel of Figure \ref{fig:beta}
shows the $\xiion$ of a variety of BC03 constant SFR models as a function of UV slope
$\beta$, over the range $-2.1\lesssim \beta \lesssim-1.7$ suggested by the 
\citet{dunlop2012b} color measurements and further constrained such that the
age of the stellar populations is less than the age of the universe at
redshift $z\sim7$ ($t\approx7.8\times10^{8}~\mathrm{yr}$).
We will limit our further discussion of BC03 models to constant SFR histories, as
we find single bursts populations display too wide a range of $\xiion$ with $\beta$
to be tightly constrained by $\beta$ alone (although limits on the luminosity will
constrain the available $\xiion$ for a given $\beta$ for single burst models, without 
the individual SED fits to objects we cannot usefully constrain such models over the
wide range of luminosity and redshifts we examine).

Three broad types of BC03 constant SFR models are 
consistent with values of $\beta=-2$. 
Generically, the BC03 constant SFR models evolve
from large values of $\xiion$ and large negative values of $\beta$ at early times to
a roughly horizontal evolution with constant $\xiion$ with $\beta$ increasing at late
times ($t\gtrsim10^{8}~\mathrm{yr}$).
Metal poor ($Z<Z_{\sun}$) constant SFR populations without dust produce values of $\beta$
significantly bluer (more negative) than observed at $z\sim7-9$ \citep{dunlop2012b}.
Mature ($\gtrsim10^{8}~\mathrm{yr}$ old), metal-rich ($Z\sim Z_{\sun}$), dust free stellar populations
evolve to a constant value of $\log \xiion \approx 24.95-25.2~\log~\xiionunit$ over the observed
$\beta$ values (as Figure \ref{fig:beta} indicates, the results are largely independent of 
IMF for constant SFR models).
Applying a dust absorption of $\AV\sim0.1$ 
using the model of \citet[][with parameters $\tauV=0.25$ and an ISM attenuation
fraction of 0.3]{charlot2000a}
shifts the SED evolution tracks down in $\xiion$ (from dust absorption) and to redder $\beta$, such that
young, metal-rich stellar populations with dust can also reproduce values of $\beta\approx-2$ while maintaining
$\log \xiion = 24.75-25.35~\log~\xiionunit$.  Moderately metal-poor models ($Z\sim0.2-0.4Z_{\sun}$) 
with as much as $\AV\approx0.1$ can also reproduce 
$\beta\approx-2$ for population ages $t>10^{8}~\mathrm{yr}$, but the most metal poor models in this range require
$t>4\times10^{8}~\mathrm{yr}$ (an initial formation redshift of $z\gtrsim12$ if observed at $z\sim7$).
We note that our conclusions about the connection between $\beta$ (or $\JAWFC-\HWFC$ color) 
and the properties of stellar populations
is wholly consistent with previous results in the literature \citep[e.g., Figure 7 of ][]{finkelstein2010a}.

While the $\beta$ measurements of \citet{dunlop2012b} have greatly reduced the available
BC03 stellar population model parameter space, there is still a broad allowable range of 
$\log~\xiion \approx 24.75-25.35~\log~\xiionunit$ available for constant SFR models with UV
spectral slopes of $\beta\approx-2$.  We therefore adopt the value
\begin{equation}
\log~\xiion = 25.2~\log~\xiionunit~~(\mathrm{adopted}) 
\end{equation}
\noindent
throughout the rest of the paper.  This $\xiion$ is in the upper range of the available values
shown in Figure \ref{fig:beta}, but is comparable to values adopted elsewhere in the
literature \citep[e.g.,][ who assume $\log \xiion \approx 25.3$ for $\beta=-2$, see their Equations 5 and 6]{kuhlen2012a}.

We also considered stellar populations reddened by nebular continuum emission 
\citep[e.g.,][]{schaerer2009a,schaerer2010a,ono2010a,robertson2010a}, which in principle could allow
relatively young, metal poor populations with larger $\xiion$ to fall into the window of $\beta$ values found by 
\citet{dunlop2012b}.  We find that for $\fesc\sim0.2$, nebular models applied to 
young ($<100~\mathrm{Myr}$) constant star formation rate BC03 models 
are still marginally too blue 
($\beta\sim-2.3$). 
Although more detailed modeling is always possible to explore the impact
of nebular emission on $\xiion$, the uniformity observed
by \citet{dunlop2012b} in the average value of $\beta$ over a range in galaxy
luminosities may argue against a diverse mixture
young and mature stellar populations in the current $z\simeq7-8$ samples.
However, as \citet{dunlop2012b} noted, a larger intrinsic scatter could be
present in the UV slope distribution of the observed population but not yet detected.
Similarly, top heavy initial mass function stellar populations with low metallicity, like the
$1-100\Msun$ Salpeter IMF models of \citet[][]{schaerer2003a} used by \citet{bouwens2010a}
to explain the earlier HUDF09 data, are disfavored owing to their blue spectral slopes.

For reference, for conversion from UV luminosity spectral density to SFR we note that for population 
ages $t>10^{8}~\mathrm{yr}$
a constant SFR BC03 model with a \citet{chabrier2003a} IMF and solar metallicity provides a $1500\mathring{A}$ luminosity spectral
density of
\begin{equation}
\label{eqn:luv_to_sfr}
L_{\mathrm{UV}} \approx 1.25 \times 10^{28} \times \frac{\mathrm{SFR}}{M_{\sun}~\mathrm{yr}^{-1}}~\LUVunit,
\end{equation}
while, as noted by \citet{madau1998a}, a comparable \citet{salpeter1955a} model provides 64\% of this UV luminosity. 
A very metal-poor population ($Z=Z_{\sun}/200$) would provide 40\% more UV luminosity per unit SFR.

%
%
\section{Ultraviolet Luminosity Density}
\label{section:rho_uv}

In addition to constraints on the spectral energy distributions of
high-redshift galaxies \citep{dunlop2012b}, the UDF12 observations
provide a critical determination of the luminosity function of star forming
galaxies at redshifts $7\lesssim z \lesssim 9$.  As described in
Section \ref{section:reionization}, when calculating the comoving
production rate $\dniondt$ of hydrogen ionizing photons per unit volume 
(Equation \ref{eqn:dniondt}) the UV luminosity density $\rhoUV$ provided
by an integral of the galaxy luminosity function is required 
(Equation \ref{eqn:rhoUV}).  An accurate estimate of the $\rhoUV$ 
provided by galaxies down to observed limits requires a careful 
analysis of star-forming galaxy samples at faint magnitudes.
Using the UDF12 data, \citet{schenker2012a} and \citet{mclure2012a}
have produced separate estimates of the $z\sim7-8$ galaxy luminosity
function for different sample selections (color-selected drop-out and
spectral energy distribution-fitted samples, respectively). As we
demonstrate, the UV luminosity densities computed from these
separate luminosity functions are consistent within $1-\sigma$ at
$z\sim7$ and in even closer agreement at $z\sim8$. Further,
\citet{mclure2012a} have provided the first luminosity function 
estimates at $z\sim9$.  Combined, these star-forming galaxy luminosity
function determination provide the required constraints on $\rhoUV$
in the epoch $z\gtrsim7$ when, as we show below, the ionization
fraction of the IGM is likely changing rapidly.

\begin{figure*}
\figurenum{2}
\epsscale{1.1}
\plotone{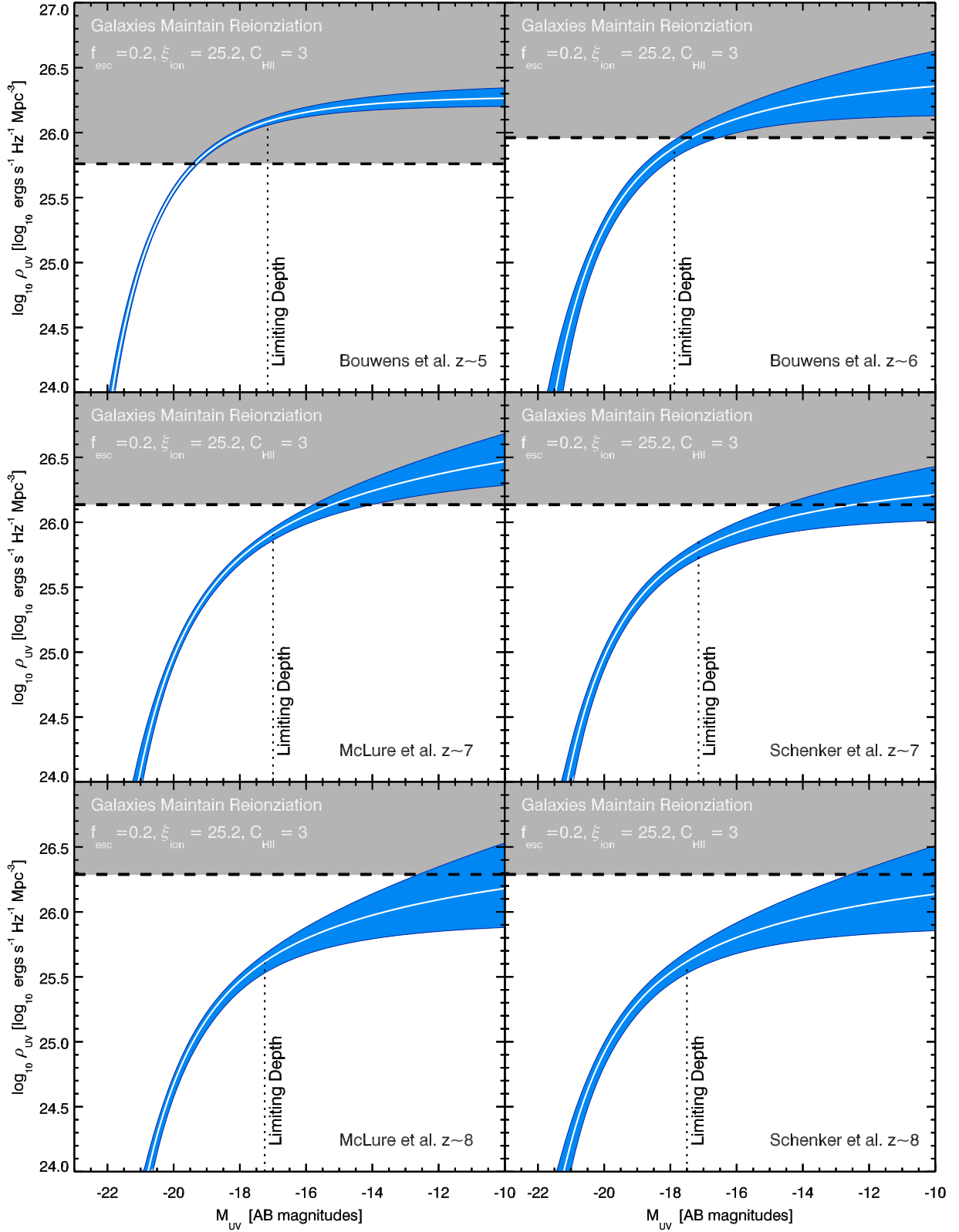}
\caption{\label{fig:lfs}
Constrained ultraviolet (UV) luminosity densities $\rhoUV$ as a function of limiting magnitude $\MUV$ and
redshift $z$.  
Shown are the $z\sim5-8$ maximum likelihood values of $\rhoUV$ vs. limiting magnitude
calculated using Equation \ref{eqn:rhoUV} (white lines), and the corresponding marginalized 
inner 68\% credibility intervals for $\rhoUV(\MUV)$ (blue regions).  In each panel, we indicate with a 
dotted line the limiting depth of the luminosity function determinations.
Also shown is the $\rhoUV$ 
required for galaxies to maintain a fully ionized universe assuming $\log \xiion = 25.2~\log \xiionunit$,
$\fesc=0.2$, $\CHII=3$, and case A recombination (dashed lines and grey regions).
We use Bayesian parameter estimation methods to determine the \citet{schechter1976a} function
parameter posterior distributions inferred from the stepwise maximum likelihood luminosity function (LF) data 
of \citet{bouwens2007a} at $z\sim4-6$ and
\citet{mclure2012a} at $z\sim7-8$.  
We also use the full posterior distribution sampling of the \citet{schechter1976a} function
parameters from the LF determination of \citet{schenker2012a} to produce additional, independent 
constraints on $\rhoUV$ at $z\sim7-8$.  
At $z\sim9$ where the data is limited, we simply infer the
LF normalization $\phistar$ keeping the characteristic magnitude $\Mstar$ and faint-end slope $\alpha$ 
fixed at the $z\sim8$ values determined by \citet{mclure2012a} and expect the inferred possible 
variation in $\rhoUV(z\sim9)$ 
to be somewhat underestimated. 
}
\end{figure*}

Given the challenge of working at the limits of the observational
capabilities of {\it HST} and the relatively small volumes probed
by the UDF (with expected cosmic variance of $\sim30\% -40\%$
at redshifts $z\sim7-9$, see \citealt{robertson2010b,robertson2010c,munoz2010a},
and Section 4.2.3 of \citealt{schenker2012a}), we anchor our constraints
on the evolving UV luminosity density with precision determinations
of the galaxy luminosity function at redshifts $4\lesssim z \lesssim 6$ by
\citet{bouwens2007a}.

To utilize as much information as possible about the luminosity function (LF)
constraints at $z\sim4-9$, we perform Bayesian inference to generate 
full posterior distributions of the galaxy luminosity functions at each
redshift.  To achieve this, we perform \citet{schechter1976a} function
parameter estimation at $z\sim4-9$ using the stepwise maximum likelihood
UV luminosity function constraints reported in Table 5 of 
\citet[][for $z\sim4-6$]{bouwens2007a} and Table 2 of 
\citet[][for $z\sim7-8$]{mclure2012a} allowing all parameters (the
luminosity function normalization $\phistar$, the characteristic
galaxy magnitude $\Mstar$, and the faint-end slope $\alpha$) to vary.
For these LF determinations, we assume Gaussian errors and a $\chi^{2}$
likelihood.
For additional constraints at $z\sim7-8$, we use the samples from the
full posterior distributions of LF parameters determined by 
\citet{schenker2012a}, using their method for Bayesian inference of LF 
parameters discussed in their Section 4.2. 
Lastly, at redshift $z\sim9$ we perform parameter
estimation on $\phistar$ given the stepwise maximum likelihood LF determination
provided in Table 4 of \citet{mclure2012a} while keeping $\Mstar$ and $\alpha$
fixed at the best-fit $z\sim8$ values reported by \citet{mclure2012a}. Again we
assume Gaussian errors and a $\chi^{2}$ likelihood. 
The limited information available at $z\sim9$ and our restricted fitting method
at this redshift mean that the inferred allowed variation in the $\rhoUV(z\sim9)$
will be underestimated. However, we have checked that this restriction does not
strongly influence our results presented in Section \ref{section:constraints}.
In each case our maximum likelihood luminosity function parameters are consistent
within $1-\sigma$ of 
the values originally reported by \citet{bouwens2007a} and
\citet{mclure2012a}, and are of course identical in the case of the \citet{schenker2012a}
LFs.  

Figure \ref{fig:lfs} shows the integrated UV luminosity density $\rhoUV$ as a function of
limiting magnitude for these galaxy luminosity function determinations at 
$z\sim5-8$ (constraints at $z\sim4$ and $z\sim9$ are also used).  
For each redshift, we show the maximum likelihood $\rhoUV$ (white lines) and
the inner 68\% variation in the marginalized $\rhoUV$ (blue regions). Since the
luminosity functions are steep ($\alpha\lesssim-1.7$), the luminosity 
densities $\rhoUV$ increase dramatically below the characteristic magnitude $\Mstar$
at each redshift, but especially at $z\gtrsim7$. The total $\rhoUV$ supplied by
star-forming galaxies therefore strongly depends on the limiting magnitude adopted to
limit the integral in Equation \ref{eqn:rhoUV}. 
The UDF12 campaign depth of $\MUV<-17$ provides $\rhoUV\approx10^{26}\rhoUVunit$ at $z\sim7$,
declining to $\rhoUV\sim3.2\times10^{25}~\rhoUVunit$ at $z\sim8$.
To put these $\rhoUV$ values 
in context, we also indicate the critical values of $\rhoUV$ required
to keep $\dQHIIdt=0$ in Equation \ref{eqn:QHII} and maintain reionization 
(grey areas and dashed line; assuming case A 
recombination\footnote{When calculating whether the production rate of ionizing
photons from galaxies can maintain the ionization fraction of the IGM near the
end of reionization, an assumption of case A recombination is appropriate since
recombinations to the hydrogen ground state largely do not help sustain the
IGM ionization. For a detailed discussion, see \citet{furl05-rec,faucher-giguere2009a,kuhlen2012a}. 
We therefore adopt case A recombination in Figure \ref{fig:lfs}.  However, we assume 
case B recombination throughout the rest of the paper.}) if
$\fesc=0.2$, $\log \xiion = 25.2~\log~\xiionunit$, and $\CHII=3$. Under these
reasonable assumptions, the currently observed
galaxy population clearly is not abundant enough to maintain reionization at $z\gtrsim7$.
Understanding the role of galaxies in the reionization process will therefore likely
require extrapolations to luminosities beyond even the UDF12 depth, and the constraints
on the LF shape achieved by the UDF12 observations will be important for performing these
extrapolations reliably.  The extent of the extrapolation down the luminosity function required for
matching the reionization constraints depends in detail on our assumptions for the escape fraction
$\fesc$ or the ionizing photon production rate per UV luminosity $\xiion$ of galaxies.  We emphasize
that the critical ionizing photon production rate $\dniondt$ depends on the product 
$\dniondt=\fesc\xiion\rhoUV$, and will shift up or down proportionally to $\fesc\xiion$ 
at fixed $\rhoUV$.  Similarly, the $\dniondt$ value that balances recombination is proportional
to the clumping factor $\CHII$, and variation in the clumping factor will also shift the required
$\rhoUV$ up or down.

%
%
\subsection{UV Luminosity Density Likelihoods}
\label{section:rho_uv_like}

We wish to use the evolving UV luminosity density $\rhoUV$ to constrain the availability of ionizing
photons $\dniondt$ as given in Equation \ref{eqn:dniondt}.  Given how significantly
$\rhoUV$ increases with the limiting absolute magnitude $\MUV$, we must 
evaluate the full model presented in Section \ref{section:reionization} for
chosen values of $\MUV$. We select $\MUV=-17$, $\MUV=-13$, and $\MUV=-10$ to provide a broad
range of galaxy luminosities extending to extremely faint magnitudes.  
The observations probe to $\MUV=-17$, and this magnitude therefore serves as a natural limit.
We consider limits as faint as $\MUV=-10$ as this magnitude may correspond to the minimum 
mass dark matter halo able to accrete gas from the photoheated intergalactic 
medium or the minimum mass dark matter halo able to retain its gas supply 
in the presence of supernova feedback. These scenarios have so far proven 
impossible to distinguish 
empirically \citep{munoz11}.
To make use of the
constraints of the UV luminosity density on $\dniondt$, we need to adopt a 
likelihood for use with Bayesian inference on parameterized forms of $\rhoUV(z)$.  
Given the parameterized form of the \citet{schechter1976a} function, we find that the
marginalized posterior distributions of $\rhoUV$ are skewed at $z\gtrsim6$, 
with tails extending to larger $\log \rhoUV$ values than a Gaussian approximation would
provide.  To capture this skewness, when constraining the cosmic reionization history in 
Section \ref{section:constraints} below, we
therefore use the full marginalized posterior distribution of $\rhoUV$ provided by the integrated
LF determinations calculated in Section \ref{section:rho_uv}.

\begin{figure*}
\figurenum{3}
\epsscale{1.1}
\plotone{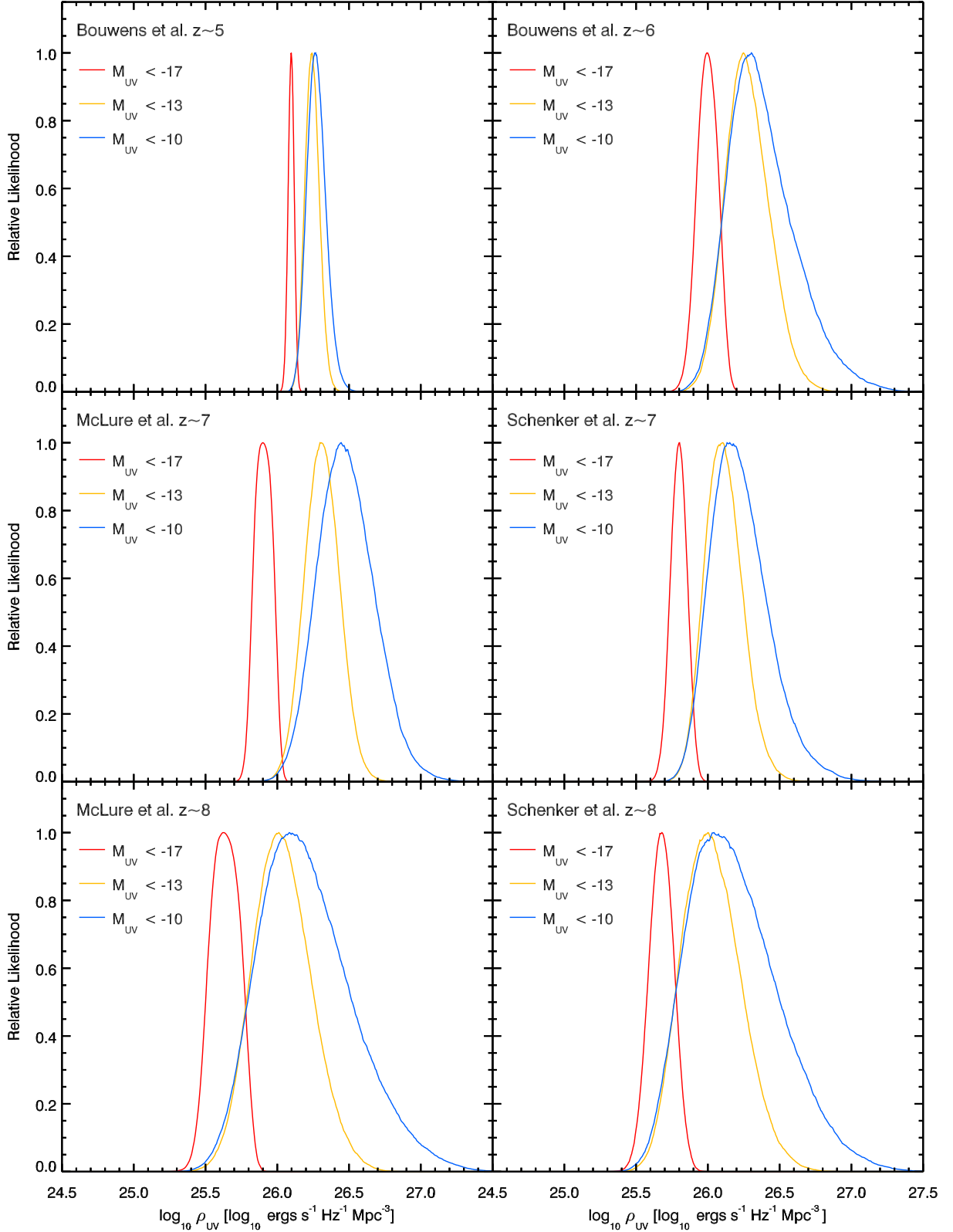}
\caption{\label{fig:likelihoods}
Likelihood functions of the UV luminosity density $\rhoUV$ for different limiting magnitudes and redshifts
used to constrain the reionization history.  Shown are the $z\sim5-8$
marginalized posterior distributions for $\rhoUV$ determined from \citet{schechter1976a} function fits to luminosity
function (LF) data, as reported in Figure \ref{fig:lfs}, for limiting magnitudes of $\MUV<-17$ (red lines), 
$\MUV<-13$ (orange lines), and $\MUV<-10$ (blue lines).  To infer the
distributions at $z\sim4-9$, we use the LF data of \citet{bouwens2007a} at $z\sim4-6$, \citet{mclure2012a} at $z\sim7-9$, and
\citet{schenker2012a} at $z\sim7-8$.  In our posterior distributions for $\rhoUV(z\sim9)$, we keep the
characteristic magnitude $\Mstar$ and faint-end-slope $\alpha$ values fixed at the $z\sim8$ best 
fit values reported by \citet{mclure2012a} but expect the width of distribution to be somewhat overly narrow.
These posterior distributions on $\rhoUV$ are used as likelihood functions when fitting a parameterized model
to the evolving UV luminosity density.  
}
\end{figure*}

Figure \ref{fig:likelihoods} shows the marginalized posterior distribution of the 
UV luminosity density $\rhoUV$ for limiting magnitudes of $\MUV=-17$ (red lines), 
$\MUV=-13$ (orange lines), and $\MUV=-10$ (blue lines)
for our Schechter function fits to the $z\sim5-6$ LFs of \citet{bouwens2007a}, the
$z\sim7-8$ LFs of \citet{schenker2012a}, and the $z\sim7-8$ LFs of \citet{mclure2012a}.
Although \citet{schenker2012a} and \citet{mclure2012a} use the same data sets, their luminosity
function determinations are based on different selection techniques.  They therefore
represent independent determinations of the high-redshift luminosity functions and are
treated accordingly.
We additionally use constraints from $z\sim4$ \citep{bouwens2007a} and $z\sim9$ \citep{mclure2012a}.
The posterior distributions for $\rhoUV(z\sim9)$ we calculate are likely
underestimated in their width owing to an assumption of fixed $\Mstar$ and $\alpha$
values, but this assumption does not strongly influence our results.
In what follows, these posterior distributions on $\rhoUV$ are used as likelihood functions 
when fitting a parameterized model to the evolving UV luminosity density.

%
%


%
%
\section{Constraints on Reionization}
\label{section:constraints}

The observational constraints on the process of cosmic reionization in the
redshift range $z\gtrsim7$ are the spectral character
of high-redshift star-forming galaxies determined from the UDF12
program (Section \ref{section:beta} and
\citealt{dunlop2012b}), the evolving luminosity density constraints enabled by
those same observations (Section \ref{section:rho_uv}, \citealt{ellis2012a}, 
\citealt{schenker2012a}, and \citealt{mclure2012a}), and the electron
scattering optical depth inferred from the 9-year {\it WMAP} observations of the
CMB (\citealt{hinshaw2012a} and briefly below).  

As we demonstrate, 
these constraints are in tension when taking the current data at face value. 
The Lyman continuum photon production
rates per unit UV luminosity spectral density calculated using the 
BC03 models that are consistent with
the UV spectral slopes of galaxies at $z\sim7-9$ are 
$\log \xiion \approx 24.8-25.3~\log~\xiionunit$ (Section \ref{section:beta}),
and for a reasonable escape fraction of $\fesc\sim0.2$ and IGM clumping factor
of $\CHII$, UV luminosity density of $\log \rhoUV> 26~\log~\rhoUVunit$ is
required to induce significant ionization at $z\ge7$, but as we showed
in \citet{ellis2012a} the
observed abundance of star-forming galaxies continues a measured decline at high-redshift ($z>8$).
Further, reproducing the {\it WMAP} Thomson optical depth requires an extended
reionization process, since instantaneous reionization ($\QHII=1$) would need to occur 
at $z=10.3\pm1.1$ to reproduce the measured $\tau=0.084\pm0.013$ \citep{hinshaw2012a}
and very likely the ionization fraction $\QHII<1$ at $z\gtrsim7$.
Based on the UDF12 and {\it WMAP} constraints, we
therefore anticipate that the UV luminosity density declines to some minimum level beyond $z>7$, and then
persists with redshift to sustain a sufficient partial ionization of the IGM to
satisfy the Thomson optical depth constraint. We need a methodology to quantify this process and, given
its redshift dependence, to determine the
required minimum luminosity of abundant star-forming galaxies to reionize the universe by $z\sim6$.

Our chosen methodology is to use a simple parameterized model of the evolving UV 
luminosity density to calculate the redshift-dependent ionizing photon production rate density
$\dniondt$, constrained to reproduce the UV luminosity density values reflected by the
likelihood functions shown in Figure \ref{fig:likelihoods}.  The evolving $\dniondt$ with
redshift is used to calculate the reionization history $\QHII(z)$ by integrating Equation \ref{eqn:QHII}.
The corresponding Thomson optical depth is calculated using Equation \ref{eqn:tau} and then 
evaluated against the posterior distribution of $\tau$ provided by the public 9-year 
{\it WMAP} Monte Carlo Markov Chains. At each posterior sample evaluation,
our method requires a full reconstruction of the reionization history and integration of the
electron scattering optical depth and we have incorporated the reionization calculation 
into the {\it MultiNest} Bayesian inference software
\citep{feroz2008a,feroz2009a}.

\subsection{A Parameterized Model for the Evolving UV Luminosity Density}
\label{section:rho_uv_model}

To infer constraints on the reionization process, we must adopt a flexible
parameterized model for the evolving UV luminosity density.  The model must
account for the decline of $\rhoUV$ apparent in Figures \ref{fig:lfs} 
and \ref{fig:likelihoods}, without artificially extending the trend in 
$\rhoUV$ at $z\lesssim6$ to redshifts $z\gtrsim8$.  The rapid decline in
$\rhoUV$ between $z\sim4$ and $z\sim5$ suggests a trend of 
$d\log\rhoUV/d\log z\sim-3$, but the higher redshift values of $\rhoUV$
flatten away from this trend, especially for faint limiting magnitudes.
Beyond $z\sim10$, we expect that some low-level UV luminosity density will
be required to reproduce the Thomson optical depth, to varying degrees
depending on the chosen limiting magnitude.

With these features in mind, we have tried a variety of parameterized
models for $\rhoUV$.  We will present constraints using a three-parameter
model given by

\begin{equation}
\label{eqn:rhoUV_model}
\rhoUV(z) = \rhoUVA \left(\frac{z}{4}\right)^{-3} + \rhoUVB \left(\frac{z}{7}\right)^{\gamma}.
\end{equation}
\noindent
The low-redshift amplitude of this model is anchored by the UV luminosity density 
at $z\sim4$, $\rhoUVA$ with units of $\rhoUVunit$. The high-redshift evolution is
determined by the normalization $\rhoUVB$ at $z\sim7$, provided in units 
of $\rhoUVunit$, and the power-law slope $\gamma$.  Using this model, we compute
the evolving $\rhoUV(z)$ and evaluate the parameter likelihoods as described immediately
above.

We have examined other models, including general broken (double) power-laws and low-redshift
power-laws with high-redshift constants or fixed slope power-laws.  Single power-law models
tend to be dominated by the low-redshift decline in the $\rhoUV$ and have difficulty 
reproducing the Thomson optical depth.  Generic broken power-law models have a degeneracy
between the low- and high-redshift evolution, even for fixed redshifts about which the
power laws are defined, and are disfavored based on their Bayesian information relative
to less complicated models that can also reproduce the $\rhoUV$ constraints.  The
model in Equation \ref{eqn:rhoUV_model} is therefore a good compromise between sufficient
generality and resulting parameter degeneracies.

However, when using this model, we use a prior to limit $\gamma<0$ to prevent 
{\it increasing} $\rhoUV$ beyond $z\sim9$.  The best current constraints on the
abundance of $z>8.5$ galaxies is from \citet{ellis2012a}, where we demonstrated that
the highest-redshift galaxies (with $\MUV\lesssim-19$) continue the smooth decline 
in abundance found at slightly later times.
Correspondingly, the constraint on $\gamma$ amounts to a limit on introducing a 
new population of galaxies arising at $z\gtrsim9$.
The usefulness of this
potentiality for the reionization of the universe has been noted elsewhere 
\citep[e.g.,][]{cen2003a,alvarez2012a}.  While imposing no constraint besides
$\gamma<0$, we typically find that nearly constant, low-level luminosity densities
at high-redshift are nonetheless favored (see below). Results for models that
feature low-redshift power-law declines followed by low-level constant $\rhoUV$
at high-redshift will produce similar quantitative results. All parameterized models 
we have examined that feature a declining or constant $\rhoUV$ and can reproduce
the Thomson optical depth constraint produce similar results, and we conclude
that our choice of the exact form of Equation \ref{eqn:rhoUV_model} is not critical.

\begin{figure*}
\figurenum{4}
\epsscale{1.2}
\plotone{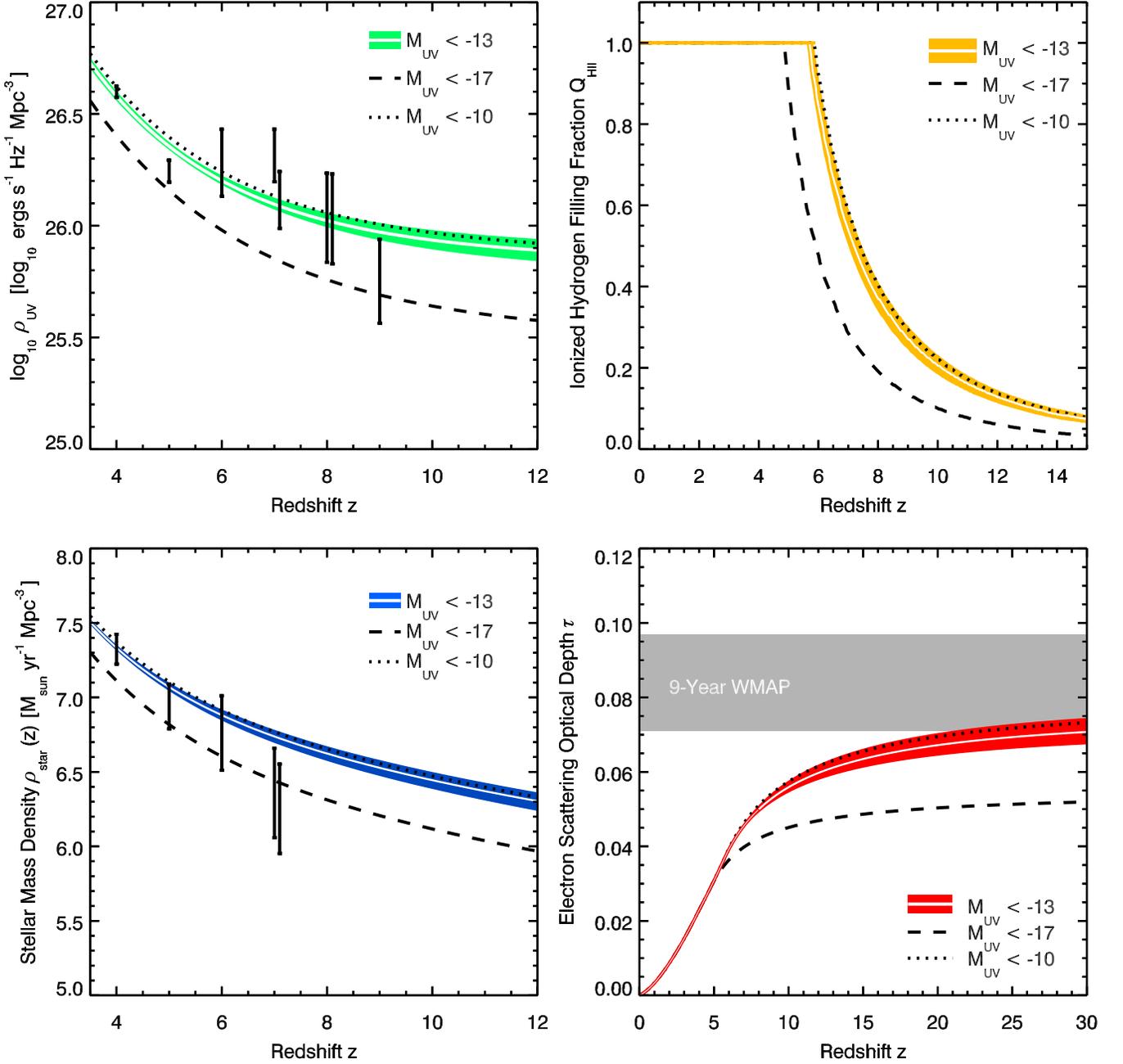}
\caption{\label{fig:summary}
Joint constraints on the reionization history, assuming ionizing photon 
contributions from galaxies with $\MUV<-17$ (maximum likelihood model shown as
dashed line in all panels), $\MUV<-10$ (maximum likelihood model shown as dotted
line in all panels), and $\MUV<-13$ (maximum likelihood model shown as white line
in all panels; 68\% credibility regions shown as colored areas).
We use the posterior distributions 
for $\rhoUV$ with redshift shown in Figure \ref{fig:likelihoods}, extrapolated
stellar mass density constraints, and the
posterior distribution on the electron scattering optical depth \citep{hinshaw2012a}
as likelihood functions to constrain the simple parameterized model for the evolving
UV luminosity density given by Equation \ref{eqn:rhoUV_model} at redshifts $z\gtrsim4$.
The constrained evolution of $\rhoUV$ is shown in the upper left panel (error bars
indicate the $\rhoUV$ constraints for $\MUV<-13$, but each model uses the appropriate
constraints).
From
$\rhoUV(z)$, we can simply integrate with redshift to determine the stellar mass densities 
(bottom left panel; data points with error bars indicate 
extrapolations of the \citet{stark2012a} stellar mass
densities to $\MUV<-13$, but all models use the appropriate constraints).  The models
tend to exceed slightly the stellar mass densities at the highest-redshifts ($z\sim7$),
a result driven by the constraint on the election scattering optical depth.
By assuming the well-motivated values of the ratio of Lyman continuum photon
production rate to UV luminosity $\log \xiion = 25.2~\log \xiionunit$ for individual sources,
an ionizing photon escape fraction $\fesc=0.2$, and an intergalactic medium clumping factor of
$\CHII=3$, the reionization history $\QHII$ calculated by integrating Equation \ref{eqn:QHII}
is shown in the upper right panel.  Integrating the reionization history provides the
electron scattering optical depth (lower right panel, 9-year {\it WMAP} constraint indicated as the
grey region).
}
\end{figure*}

\subsection{Reionization Constraints from Galaxies}
\label{section:reion-gal}

Figure \ref{fig:summary} shows 
constraints on the reionization process calculated using the model
described in Section \ref{section:reionization}.  
We perform our Bayesian
inference modeling assuming $\MUV<-17$ (close to the UDF12 limit 
at $z\sim8$, maximum likelihood model is shown as a dashed line 
in all panels), $\MUV<-10$ (an extremely
faint limit, with the maximum likelihood model shown as a  
dotted line in all panels), and an intermediate limit
$\MUV<-13$ (colored regions show 68\% credibility regions, while
white lines indicate the maximum likelihood model).  We now will
discuss the UV luminosity density, stellar mass density, ionized
filling fraction, and electron scattering optical depth results in turn.

The upper left hand panel shows parameterized models of the
UV luminosity density (as given in Equation \ref{eqn:rhoUV_model}),
constrained by observations of the UV luminosity density (inferred from
the measured luminosity functions, see
Figures \ref{fig:lfs} and \ref{fig:likelihoods}) integrated down to 
each limiting magnitude.  The figure shows error bars to indicate the
68\% credibility width of the posterior distributions of $\rhoUV$ at redshifts $z\sim4-9$
integrated to $\MUV<-13$, but the likelihood of each model is calculated
using the full marginalized $\rhoUV$ posterior distributions 
appropriate for its limiting magnitude.
In each case the $\rhoUV(z)$ evolution matches well the constraints
provided by the luminosity function extrapolations, which owes to the
well-defined progression of declining $\rhoUV$ with redshift inferred
from the UDF12 and earlier datasets.  
For reference, the maximum likelihood values for the parameters of Equation
\ref{eqn:rhoUV_model} for $\MUV<-13$ are $\log~\rhoUVA = 26.50~\log~\rhoUVunit$, $\log~\rhoUVB = 25.82~\log~\rhoUVunit$,
and $\gamma = -0.003$.

What drives the constraints on $\rhoUV$ for these limiting magnitudes?  
Some tension
exists at high-redshift, as the models prefer a relatively flat $\rhoUV(z)$ 
beyond the current reach of the data.
In each case, the 
maximum likelihood models become flat at high-redshift and reflect the
need for continued star formation at high-redshift to sustain a low-level
of partial IGM ionization.
To better answer this question, 
we have to calculate the full stellar mass density evolutions, reionization histories, 
and Thomson optical depths of each model.

The lower left hand panel shows how the models compare to the stellar mass densities
extrapolated from the results by \citet{stark2012a}, using the method
described in Section \ref{section:reionization}.  The error bars in the figure reflect
the stellar mass densities extrapolated for contributions from galaxies with $\MUV<-13$,
but the appropriately extrapolated mass densities are used for each model.  The error
bars reflect bootstrap-calculated uncertainties that account for statistically possible 
variations
in the best fit stellar mass - UV luminosity relation given by Equation \ref{eqn:Mstar_MUV},
and are $\sigma\approx0.3$~dex at $z\sim7$.  To balance the relative constraint from the stellar
mass density evolution with the single optical depth constraint (below), we assign the likelihood
contributions of the stellar mass density and Thomson optical depth equal weight.
  The
stellar mass densities calculated from the models evolve near the $1-\sigma$
upper limits of the extrapolated constraints, again reflecting the need for continued star formation
to sustain a low-level of IGM ionization.\footnote{We note that if we ignore the electron scattering
optical depth constraint, the other datasets favor a UV luminosity density that declines more rapidly
at high redshift than does our maximum likelihood model calculated including all the constraints.
However, the UV luminosity
density parameters in Equation \ref{eqn:rhoUV_model} remain similar. Excluding the stellar mass
density constraint has almost no effect on the maximum likelihood parameters inferred for the model.}  

Assuming a ratio of Lyman continuum photon
production rate to UV luminosity $\log \xiion = 25.2~\log \xiionunit$ for individual sources
consistent with UV slopes measured from the UDF12 data \citep{dunlop2012b},
an ionizing photon escape fraction $\fesc=0.2$, and an intergalactic medium clumping factor of
$\CHII=3$, the reionization history $\QHII(z)$ produced by the evolving $\rhoUV$ can be
calculated by integrating Equation \ref{eqn:QHII} and is shown in the upper right panel of
Figure \ref{fig:summary}.
We find that the currently observed galaxy population at magnitudes brighter than $\MUV<-17$ 
(dashed line) can only manage to reionize fully the universe at late times $z\sim5$, with 
the IGM 50\% ionized by $z\sim6$ and $<5\%$ ionized at $z\sim12$.  Contributions from galaxies
with $\MUV<-13$ reionize the universe just after $z\sim6$, in agreement with a host of additional
constraints on the evolving ionized fraction (see Section \ref{section:comp-reion} below) and
as suggested by some previous analyses \citep[e.g.,][]{salvaterra2011a}.  These
galaxies can sustain a $\sim50\%$ ionized fraction at $z\sim7.5$, continuing to $\sim12-15\%$
at redshift $z\sim12$.
Integrating
further down to $\MUV<-10$ produces maximum likelihood models that reionize the universe slightly earlier.

The Thomson optical depths resulting from the reionization histories can be determined by
evaluating Equation \ref{eqn:tau}.  
To reproduce the 9-Year {\it WMAP} $\tau$  values, 
most models display low levels of UV luminosity density that persist to high redshift.
Even maintaining a flat $\rhoUV$ at the maximum level allowed
by the luminosity density constraints at redshifts $z\lesssim9$, as the maximum likelihood models do,
the optical depth just can 
be reproduced.  When considering only the currently observed population $\MUV<-17$, the UV 
luminosity density
would need to increase toward higher $z>10$ redshifts for the universe to be fully reionized
by redshift $z\sim6$ and the Thomson optical depth to be reproduced.

We remind the reader that while the exact results for, e.g., the $\QHII$ evolution of course depends on the
choices for the escape fraction $\fesc$ or the ionizing photon production rate $\xiion$. If $\fesc$
or $\xiion$ are lowered, the evolution of $\QHII$ is shifted toward lower redshift.  For instance, with
all other assumptions fixed, we find that complete reionization is shifted to $z\sim5$ for $\fesc=0.1$
and $z\sim4.75$ for $\log~\xiion=25.0~\log~\xiionunit$.  Complete reionization can also be made earlier by
choosing a smaller $\CHII$ or delayed by choosing a larger $\CHII$.

Given the above results, we conclude that under our assumptions (e.g., $\fesc=0.2$ and $\CHII\approx3$)
galaxies currently observed down to the limiting magnitudes of deep high-redshift surveys (e.g., UDF12) do not
reionize the universe alone,
and to simultaneously reproduce the $\rhoUV$ constraints, the $\tau$ constraint,
and to reionize the universe by $z\sim6$ requires yet fainter populations.
This
conclusion is a ramification of the UDF12 UV spectral slope constraints by \citet{dunlop2012b},
which eliminate the possibility that increased Lyman continuum emission by metal-poor populations
could have produced reionization at $z>6$ \citep[e.g.,][]{robertson2010a}.
However, too much additional
star formation beyond the $\MUV<-13$ models shown in Figure \ref{fig:summary} will begin to exceed 
the stellar mass density constraints, depending on $\fesc$.  
It is therefore interesting to know whether the $\MUV<-13$ models
that satisfy the $\rhoUV$, $\rhostar$, and $\tau$ constraints also satisfy other external constraints on the
reionization process, and we now turn to such an analysis.

%
%
\section{Comparison to Other Probes of The Ionized Fraction}
\label{section:comp-reion}

\begin{figure}
\figurenum{5}
\epsscale{1.2}
\plotone{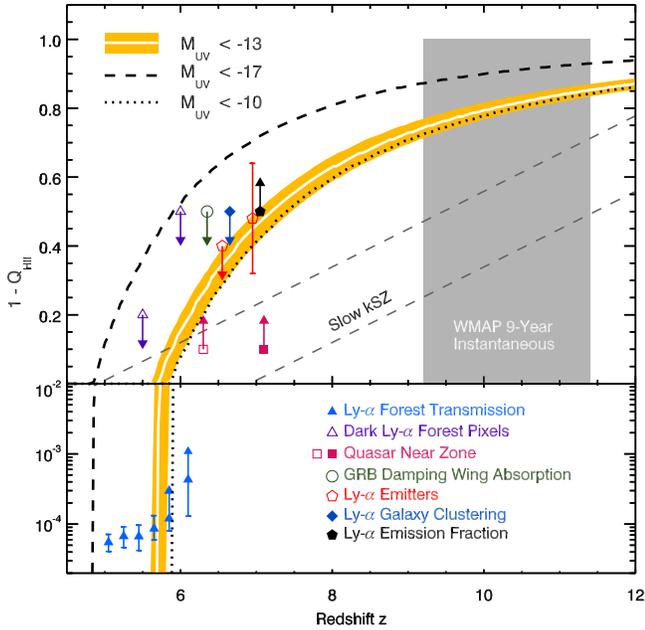}
\caption{\label{fig:additional_constraints}
Reionization histories for models that include galaxies with 
$\MUV < -13$ (maximum likelihood model: white line; 68\% credibility region: orange area), 
$\MUV < -17$ (maximum likelihood model only, dashed line), and $\MUV < -10$ (maximum likelihood model only, dotted line).
We also show a collection of other claimed 
constraints on the neutral fraction $1-Q_{\rm HII}$; note how the ordinate switches from 
log to linear to more clearly illustrate the full range of measurements. These include 
(see text for details and references): measurements of the Lyman-$\alpha$ forest transmission 
(blue solid triangles), conservative upper limits on the neutral fraction from the fraction of 
dark pixels in the Lyman-$\alpha$ forest (purple open triangles), quasar near-zone measurements 
(open and solid magenta squares), damping wing absorption in a GRB (open green circle), the evolving 
abundance of Lyman-$\alpha$ emitter galaxies (open red pentagons), the clustering of those 
galaxies (filled dark blue diamond), and the evolving fraction of Lyman-break galaxies with strong 
Lyman-$\alpha$ emission lines (filled black pentagon). The grey dashed lines labeled ``Slow kSZ'' are 
purely illustrative, showing the slowest evolution currently allowed by the small-scale CMB 
temperature data; the endpoint of reionization is arbitrary in these cases. Finally, the 
shaded gray region shows the redshift during which instantaneous reionization would occur 
according to the WMAP 9-year data.
}
\end{figure}

In this section we will collect constraints on the IGM neutral fraction from the 
literature and compare them to the evolution of this quantity in our models based 
on the UDF12 data. These constraints come from a wide variety of astrophysical 
measurements, but all are subject to substantial systematic or modeling uncertainties, 
about which we will comment below.  We show the full set in Figure~\ref{fig:additional_constraints}, 
along with a comparison to our model histories.  We will first briefly discuss these constraints and then 
how our model histories fare in comparison to them.

\subsection{The Lyman-$\alpha$ Forest} 
\label{section:forest-reion}

The best known Lyman-$\alpha$ forest constraints come from \citet{fan2006a}, who measured the effective optical depth evolution along lines of sight taken from SDSS (including both Lyman-$\alpha$ and higher-order transitions, where available).  Those authors then made assumptions about the IGM temperature and the distribution of density inhomogeneities throughout the universe to infer the evolution of the neutral fraction (these assumptions are necessary for comparing the higher-order transitions to Lyman-$\alpha$ as well, because those transitions sample different parts of the IGM).  The corresponding limits on the neutral fraction according to their model are shown by the filled triangles in Figure~\ref{fig:additional_constraints}.

The original data here are the transmission measurements in the different transitions.  
Transforming those into constraints on the ionized fraction requires a model that: 
(1) predicts the temperature evolution of the IGM \citep{hui03, trac08, furl09-igmtemp}; 
(2) describes the distribution of gas densities in the IGM \citep{miralda00, bolton09-density}; 
(3) accounts for spatial structure in the averaged transmission measurements \citep{lidz06}; 
and (4) predicts the topology of ionized and neutral regions at the tail end of reionization
\citep{fzh04, choudhury09}. 
These are all difficult, and the \citet{fan2006a} constraints -- based upon a simple semi-analytic model for the IGM structure -- can be evaded in a number of ways. In particular, their model explicitly ignores the possibility that $\QHII<1$, working only with the residual neutral gas inside a mostly ionized medium.

Unsurprisingly, our models do not match these Lyman-$\alpha$ forest measurements very well. 
The crude approach of Equation \ref{eqn:QHII} fails to address effects from the detailed density
structure of the
IGM on the evolution of the mean ionization fraction near the end of the reionization process.  Our 
model
is therefore not sufficient to model adequately the tail end of reionization, and only more
comprehensive models that account for both radiative transfer effects and the details of
IGM structure will realistically model these data at $z < 6$.

More useful to us is a (nearly) model-independent \emph{upper} limit on the neutral fraction provided by simply counting the dark pixels in the spectra \citep{mesinger10}.  \citet{mcgreer11} present several different sets of constraints, depending on how one defines ``dark" and whether one uses a small number of very deep spectra (with a clearer meaning to the dark pixels but more cosmic variance) versus a larger set of shallower spectra.  Their strongest constraints are roughly $1-\QHII < 0.2$ at $z = 5.5$ and $1-\QHII < 0.5$ at $z = 6$, 
shown by the open triangles in Figure~\ref{fig:additional_constraints}. Interestingly, this model-independent approach permits rather late reionization.

\subsection{Ionizing Background}
\label{section:ionizing_background}

Another use of the Lyman-$\alpha$ forest is to measure the ionizing background and 
thereby constrain the emissivity $\epsilon$ of galaxies: the ionization rate 
$\Gamma \propto \epsilon \lambda$, where $\lambda$ is the Lyman continuum mean free path.  
\citet{bolton2007a} attempted such a measurement at $z > 5$. 
The method is difficult as it involves interpretation of spectra near the 
saturation limit of the forest.
Nonetheless, the \citet{bolton2007a} 
analysis shows that the ionizing background falls by about a factor of ten from $z\sim3$ to 
to $z \sim 6$ \citep[though see][]{mcquinn2011a}. The resulting comoving emissivity is 
roughly the same as 
at $z \sim 2$, corresponding to 1.5--3 photons per hydrogen atom over the age of the Universe 
\citep[see also][]{haardt12}.  

Additional detailed constraints were derived by \citet{faucher-giguere2008a}, who used
quasar spectra at $2\lesssim z \lesssim 4$ to infer a photoionization rate for intergalactic
hydrogen.  Combining these measurements with the previous work by \citet{bolton2007a} and
estimates of the intergalactic mean free path of Lyman continuum photons 
\citep{prochaska2009a,songaila2010a}, \citet{kuhlen2012a} calculated constraints on the
comoving ionizing photon production rate.  
These inferred values of 
$\log~\dniondt < 51~\log~\mathrm{s}^{-1}~\mathrm{Mpc}^{-3}$ at $z\sim2-6$ 
are lower than those produced by naively extrapolating typical models of the evolving 
high-redshift UV luminosity density that satisfy reionization constraints.
To satisfy both the low-redshift IGM emissivity constraints and the
reionization era constraints then available, \citet{kuhlen2012a} posited an evolving escape
fraction 
\begin{equation}
\label{eqn:fesc_z}
\fesc(z) = f_{0} \times [(1+z)/5]^{\kappa}.
\end{equation}
If the power-law slope $\kappa$
is large enough, the IGM emissivity constraints at $z<6$ can be satisfied with a low $f_{0}$
while still reionizing the universe by $z\sim6$ and matching previous {\it WMAP} constraints on
the electron scattering optical depth.  Other previous analyses have emphasized similar needs for an
evolving escape fraction \citep[e.g.,][]{ferrara2012a,mitra2013a}.

To study the effects of including the IGM emissivity constraints by permitting an evolving $\fesc$,
we repeat our calculations in Section \ref{section:constraints} additionally allowing an escape fraction
given by Equation \ref{eqn:fesc_z} with varying $f_{0}$ and $\kappa$ (and maximum escape fraction $\fmax=1$).
With a limiting magnitude $\MUV<-13$, and utilizing the updated constraints from UDF12 and {\it WMAP}, 
we find maximum likelihood values $f_0=0.054$ and $\kappa=2.4$ entirely consistent with the results
of \citet[][see their Figure 7]{kuhlen2012a}.  
When allowing for this evolving $\fesc$ 
(with $\fmax=1$), we sensibly find that the need for low-level star formation to satisfy
reionization constraints is reduced (the maximum likelihood model has a high-redshift 
luminosity density evolution of $\rhoUV\propto z^{-1.2}$) and the agreement with the stellar mass
density constraints is improved.  

However, the impact of the evolving $\fesc(z)$ depends 
on the maximum allowed $\fesc$.  We allow only $\fesc$ to vary (not the product $\fesc\xiion$)
owing to our UV spectral slope constraints, and the maximum $\fesc=1$ (and, correspondingly, the
maximum ionizing photon production rate per galaxy) is reached by $z\sim15$.  In this model
with maximum $\fmax=1$, the escape fraction at $z\sim7-8$ (where the ionized volume filling factor
$\QHII$ is changing rapidly) is $\fesc\sim0.17-0.22$, very similar to our fiducial constant
$\fesc=0.2$ adopted in Section \ref{section:constraints}.  The rate of escaping ionizing photons
per galaxy in this evolving $\fesc$ model with maximum $\fmax=1$ trades off against the 
high-redshift UV luminosity density evolution.  
The need for a low-level of star formation out to high redshift is somewhat reduced as the
increasing escape fraction with redshift compensates to maintain a partial IGM ionization 
fraction and recover the observed {\it WMAP} $\tau$. Without placing additional constraints
on the end of reionization, this evolving $\fesc$ model with $\fmax=1$ produces a broader
redshift range of $z\sim4.5-6.5$ for the completion of the reionization process than the
constant $\fesc$ model described in Section \ref{section:constraints}.

If we instead adopt an evolving $\fesc$ model with $\fmax=0.2$ we 
recover $\rhoUV$, $\rhostar$, and Thomson optical depth evolutions that are
similar to the constant $\fesc$ model from Section \ref{section:constraints},
but can also satisfy the additional low-redshift IGM emissivity constraints.  
Compared with the comoving ionizing photon production rates 
$\dot{n} = [3.2, 3.5, 4.3]\times 10^{50}~\mathrm{s}^{-1}~\mathrm{Mpc}^{-3}$ at redshifts
$z=[4.0,4.2,5.0]$ inferred by \citet{kuhlen2012a} from measurements by 
\citet{faucher-giguere2008a}, \citet{prochaska2009a}, and \citet{songaila2010a},
this maximum likelihood model with evolving $\fesc$ recovers
similar values ($\dot{n} = [3.3, 3.5, 4.7]\times 10^{50}~\mathrm{s}^{-1}~\mathrm{Mpc}^{-3}$ at $z=[4, 4.2, 5]$).
In contrast, the constant $\fesc$ model in Section \ref{section:constraints} calculated without
consideration of these IGM emissivity constraints would produce $\dot{n}\sim8-13\times 10^{50}~\mathrm{s}^{-1}~\mathrm{Mpc}^{-3}$ at $z\sim4-5$.
The reduction of the escape fraction at $z\sim4-6$ compared with the baseline model with 
constant $\fesc$ allows full ionization of the IGM to occur slightly later in the
evolving $\fesc$ model (at redshifts as low as $z\approx5.3$, although in the maximum likelihood
model reionization completes within $\Delta z\approx0.1$ of the redshift suggested by the
maximum likelihood constant $\fesc$ model 
of Section \ref{section:constraints}).

With a similar range of
assumptions to \citet{kuhlen2012a} for a time-dependence in the escape fraction,
the model presented in Section \ref{section:constraints} can
therefore
also satisfy the low-redshift IGM emissivity constraints without assuming an escape fraction of
$\fesc\sim1$ at high redshifts.  While the evolving covering fraction in galaxy spectra \citep{jones2012a} 
provides additional
observational support for a possible evolution in the
escape fraction at $z\lesssim5$, such empirical support for evolving $\fesc$ does not yet exist at $z\gtrsim5$.

\subsection{The Lyman-$\alpha$ Damping Wing}
\label{section:damping-wing-reion}

Another use of the Lyman-$\alpha$ line is to measure precisely the shape of the red damping wing of the line: the IGM absorption is so optically thick that the shape of this red absorption wing depends upon the mean neutral fraction in the IGM \citep{miralda98}, though the interpretation depends upon the morphology of the reionization process, leading to large intrinsic scatter and biases 
\citep{mcquinn08-damp,mesinger08-damp}.  Such an experiment requires a deep spectrum of a bright source in order to identify the damping wing.

One possibility is a gamma-ray burst, which has an intrinsic power-law spectrum, making it relatively easy to map the shape of a damping wing.  The disadvantage of these sources is that (at lower redshifts) they almost always have damped Lyman-$\alpha$ (DLA) absorption from the host galaxy, which must be disentangled from any IGM signal \citep{chen07}.

To date, the best example of such a source is GRB050904 at $z=6.3$, which received rapid followup and produced a high signal-to-noise spectrum \citep{totani06}.  \citet{mcquinn08-damp} studied this spectrum in light of patchy reionization models. Because it has intrinsic DLA absorption, the constraints are relatively weak: they disfavor a fully neutral IGM but allow $\QHII\sim 0.5$. We show this measurement with the open circle in Figure~\ref{fig:additional_constraints}.

Quasars provide a second possible set of sources. These are much easier to find but suffer from complicated intrinsic spectra near the Lyman-$\alpha$ line. \citet{schroeder12} have modeled the spectra of three SDSS quasars in the range $z=6.24$--6.42 and found that all three are best fit if a damping wing is present (see also \citealt{mesinger04}). Although the spectra themselves cannot distinguish an IGM damping wing from a DLA, they argue that the latter should be sufficiently rare that the IGM must have $\QHII \la 0.1$ at 95\% confidence. We show this point as the open square in Figure~\ref{fig:additional_constraints}. This conclusion is predicated on accurate modeling of the morphology of reionization around quasars and the distribution of strong IGM absorbers at the end of reionization.

\subsection{The Near Zones of Bright Quasars}
\label{section:qso-near-zone}

Any ionizing source that turns on inside a mostly neutral medium will carve out an H~II region whose extent depends upon the total fluence of ionizing photons from the source.  If one can measure (or guess) this fluence, the extent of the ionized bubble then offers a constraint on the original neutral fraction of the medium.
This is, of course, a difficult proposition, as the extremely large Lyman-$\alpha$ IGM optical depth implies that the spectrum may go dark even if the region is still highly-ionized:  in this case, you find only a lower limit to the size of the near zone \citep{bolton07-prox}.

\citet{carilli10} examined the trends of near-zone sizes in SDSS quasars from $z=5.8$--$6.4$. The sample shows a clear trend of decreasing size with increasing redshift (after compensation for varying luminosities) by about a factor of two over that redshift range.  Under the assumption that near zones correspond to ionized bubbles in a (partially) neutral medium, the volume $V \propto (1-\QHII)^{-1}$, or, in terms of the radius, $R_{\rm NZ}^{-3} \propto (1-\QHII)$.  In that case, a two-fold decrease in size corresponds to an order of magnitude increase in the neutral fraction.  However, if the \citet{fan2006a} Lyman-$\alpha$ forest measurements are correct, the neutral fraction at $z \sim 5.8$ is so small that the zones are very unlikely to be in this regime. In that case, the trend in sizes cannot be directly transformed into a constraint on the filling factor of ionized gas.
We therefore do not show this constraint on Figure~\ref{fig:additional_constraints}, as its interpretation is unclear.

The recently discovered $z=7.1$ quasar has a very small near-zone \citep{mortlock2011a}. \citet{bolton11} used a numerical simulation to analyze it in detail. They concluded the spectrum was consistent both with a small ionized bubble inside a region with $\QHII \la 0.1$ and with a highly-ionized medium $(1-\QHII) \sim 10^{-4}$--$10^{-3}$ if a DLA is relatively close to the quasar. They argue that the latter is relatively unlikely (occurring $\sim 5\%$ of the time in their simulations). Further support to the IGM hypothesis is lent by recent observations showing no apparent metals in the absorber, which would be unprecedented for a DLA \citep{simcoe12}. A final possibility is that the quasar is simply very young ($\la 10^6$~yr) and has not had time to carve out a large ionized region. We show this constraint, assuming that the absorption comes from the IGM, with the filled square in Figure~\ref{fig:additional_constraints}.  

\subsection{The Kinetic Sunyaev-Zel'dovich Effect}
\label{section:ksz-reion}

Recent small-scale temperature measurements have begun to constrain the contribution of patchy reionization to the kinetic Sunyaev-Zel'dovich (kSZ) effect, which is generated by CMB photons scattering off coherent large-scale velocities in the IGM. These scatterings typically cancel (because any redshift gained from scattering off of gas falling into a potential well is canceled by scattering off gas on the other side), 
but during reionization modulation by the ionization field can prevent such cancellation \citep{gruzinov98,knox98}. 
There is also a contribution from nonlinear evolution, see \citet{ostriker86}.  Both the Atacama Cosmology Telescope and the South Pole Telescope have placed upper limits on this signal \citep{dunkley11,reichardt12}.  

Because the patchiness and inhomogeneity of the process induce this signal, the kSZ signal grows so long as this patchy contribution persists.  As a result, it essentially constrains the duration of reionization.  Using SPT data, \citet{zahn12} claim a limit of $\Delta z < 7.2$ (where $\Delta z$ is the redshift difference between $\QHII = 0.2$ and $\QHII = 0.99$) at 95\% confidence.  The primary limiting factor here is a potential correlation between the thermal Sunyaev-Zel'dovich effect and the cosmic infrared background, which pollutes the kSZ signal.  If that possibility is ignored, the limit on the duration falls to $\Delta z < 4.4$. \citet{mesinger12} found similar limits from ACT data. 
They showed that the limit without a correlation is difficult to reconcile with the usual models of 
reionization.
Intensive efforts are now underway to measure the correlation.

We show two examples of the \emph{slowest} possible evolution allowed by these constraints (allowing for a correlation) with the straight dashed lines in Figure~\ref{fig:additional_constraints}. Note that the timing of these curves is entirely arbitrary (as is the shape: there is absolutely no reason to expect a linear correlation between redshift and neutral fraction).  They simply offer a rough guide to the maximum duration over which substantial patchiness can persist.

\subsection{Lyman-$\alpha$ Lines in Galaxies}
\label{section:lae-reion}

The final class of probes to be considered here relies on Lyman-$\alpha$ emission lines from galaxies (including some of those cataloged in the HUDF09 and UDF12 campaigns). As these line photons propagate from their source galaxy to the observer, they pass through the IGM and can suffer absorption if the medium has a substantial neutral fraction. This absorption is generally due to the red damping wing, as the sources most likely lie inside of ionized bubbles, so the line photons have typically redshifted out of resonance by the time they reach neutral gas.

One set of constraints come from direct (narrowband) surveys for Lyman-$\alpha$ emitting galaxies. As the IGM becomes more neutral, these emission lines should suffer more and more extinction, and so we should see a drop in the number density of objects selected in this manner \citep{santos04, furl04-lya, furl06-lya, mcquinn07-lya, mesinger08-lya}.  Such surveys have a long history (e.g., \citealt{hu02-lya, malhotra04, santos04-obs, kashikawa06}).
 To date, the most comprehensive surveys have come from the Subaru telescope. \citet{ouchi10} examined $z=6.6$ Lyman-$\alpha$ emitters and found evidence for only a slight decline in the Lyman-$\alpha$ transmission.  They estimate that $\QHII \ga 0.6$ at that time.  \citet{ota08} examined the $z=7$ window. Very few sources were detected, which may indicate a rapid decline in the population. They estimate that $\QHII \approx 0.32$--0.64.  We show these two constraints with the open pentagons in Figure~\ref{fig:additional_constraints}. 

The primary difficulty with measurements of the evolution of the Lyman-$\alpha$ emitter number density is that the overall galaxy population is also evolving, and it can be difficult to determine if evolution in the emitter number counts is due to changes in the IGM properties or in the galaxy population.  
An alternate approach is therefore to select a galaxy sample independently of the Lyman-$\alpha$ line (or at least as independently as possible) and determine how the fraction of these galaxies with strong Lyman-$\alpha$ lines evolves \citep{stark2010a}.  
Such a sample is available through broadband Lyman break searches with HST and Subaru. Several groups have performed these searches to $z \sim 8$ \citep[][]{fontana10,pentericci2011a,schenker12b, ono12-lya} and found that, although the fraction of Lyman-$\alpha$ emitters increases slightly for samples of similar UV luminosity from $z \sim 3$--$6$, there is a marked decline beyond $z \sim 6.5$ \citep[see also][]{treu12}.  This decline could still be due to evolutionary processes within the population \citep[e.g., dust content, see][]{dayal2012a}, but both the rapidity of the possible evolution and its reversal from trends now well-established at lower redshift make this unlikely. Assuming that this 
decline can be attributed to the increasing neutrality of the IGM, it requires $\QHII \la 0.5$ \citep{mcquinn07-lya, mesinger08-lya, dijkstra11}. We show this constraint with the filled pentagon in Figure~\ref{fig:additional_constraints}.

There is one more signature of reionization in the Lyman-$\alpha$ lines of galaxies.  A partially neutral IGM does not extinguish these lines uniformly: galaxies inside of very large ionized bubbles suffer little absorption, while isolated galaxies disappear even when $Q_{\rm HII}$ is large.  This manifests as a change in the apparent clustering of the galaxies, which is attractive because such a strong change is difficult to mimic with baryonic processes within and around galaxies \citep{furl06-lya, mcquinn07-lya, mesinger08-lya}.  Clustering is a much more difficult measurement than the number density, requiring a large number of sources.  It is not yet possible with the Lyman-$\alpha$ line sources at $z \sim 7$, but at $z \sim 6.6$ there is no evidence for an anomalous increase in clustering \citep{mcquinn07-lya, ouchi10}, indicating $\QHII \ga 0.5$ at that time.  We show this constraint with the filled diamond in Figure~\ref{fig:additional_constraints}.

\subsection{Comparison to Our Models}
\label{section:comp-models}

Figure~\ref{fig:additional_constraints} also shows the reionization history in our preferred model, with the associated confidence intervals, that extrapolates the observed luminosity function to $\MUV = -13$. Amazingly, this straightforward model obeys all the constraints we have listed in this section, with the exception of the \citet{fan2006a} Lyman-$\alpha$ forest measurements.  However, we remind the reader that our crude reionization model fails at very small neutral fractions because it does not properly account for the high gas clumping in dense systems near galaxies, so we do not regard this apparent disagreement as worrisome to any degree.

It is worth considering the behavior of this model in some detail. A brief examination of the data points reveals that the most interesting limits come from the Lyman-$\alpha$ lines inside of galaxies (the filled pentagon, from spectroscopic followup of Lyman-break galaxies, and the open pentagons, from direct narrowband abundance at $z \sim 6.6$ and $7$). These require relatively high neutral fractions at $z \sim 7$ in order for the IGM to affect the observed abundance significantly; the model shown here just barely satisfies the constraints. In a model in which we integrate the luminosity function to fainter magnitudes (e.g., $\MUV<-10$, dotted line), 
the ionized fraction is somewhat higher at this time and may prevent the lines from being significantly extinguished, 
while a model with the minimum luminosity closer to the observed limit (e.g., $\MUV<-17$, dashed line) 
reionizes the universe so late that the emitter population at $z \sim 6.6$ should have been measurably reduced.

Clearly, improvements in the measured abundance of strong line emitters at this epoch (perhaps with new multi-object infrared spectrographs, like LUCI on LBT or MOSFIRE on Keck, or with new widefield survey cameras, like HyperSuprimeCam) will be very important for distinguishing viable reionization histories. Equally important will be improvements in the modeling of the decline in the Lyman-$\alpha$ line emitters, as a number of factors both outside (the morphology of reionization, resonant absorption near the source galaxies, and absorption from dense IGM structures) and inside galaxies (dust absorption, emission geometry, and winds) all affect the detailed interpretation of the raw measurements (e.g., \citealt{santos04, dijkstra11, bolton12}).

In order to have a relatively high neutral fraction at $z \sim 7$, a plausible reionization history \emph{cannot} complete the process by $z \sim 6.4$ (to which the most distant Lyman-$\alpha$ forest spectrum currently extends), 
unless either: (1) the Lyman-continuum luminosity of galaxies, per unit 1500~$\mathring{A}$ luminosity, evolves rapidly over that interval, (2) the escape fraction $\fesc$ increases rapidly toward lower redshifts, or 
(3) the abundance of galaxies below $\MUV \sim -17$ evolves rapidly.  Otherwise, the UDF12 luminosity functions have sufficient precision to fix the shape of the $Q_{\rm HII}(z)$ curve over this interval (indeed, to $z \sim 8$), and as shown in Figure~\ref{fig:additional_constraints}, the slope is relatively shallow.  Indeed, in our best-fit model the universe still has $1.0-Q_{\rm HII} \sim 0.1$ at $z \sim 6$. This suggests intensifying searches for the last neutral regions in the IGM at this (relatively) accessible epoch.

Although reionization ends rather late in this best-fit model, it still satisfies the WMAP optical depth constraint (at $1-\sigma$) thanks to a long tail of low-level star formation out to very high redshifts, as suggested by \citet{ellis2012a}, although the agreement is much easier if the WMAP value turns out to be somewhat high.  
This implies that the kinetic Sunyaev-Zel'dovich signal should have a reasonably large amount of power. Formally, our best-fit model has $\Delta z \sim 5$, according to the definition of \citet{zahn12}, which is within the range that can be constrained if the thermal Sunyaev-Zel'dovich contamination can be sorted out. 
We also note that the {\it WMAP} results \citep{hinshaw2012a}
indicate that when combining the broadest array of available data sets, the best fit electron scattering optical
depth lowers by $\sim5\%$ to $\tau\approx0.08$.  

In summary, we have shown that the UDF12 measurements allow a model of star formation during the cosmic dawn that satisfies all available constraints on the galaxy populations and IGM ionization history, with reasonable extrapolation to fainter systems and \emph{no assumptions about high-redshift evolution in the parameters of star formation or UV photon production and escape}. Of course, this is not to say such evolution cannot occur (see \citealt{kuhlen2012a}, \citealt{alvarez2012a}, and our Section \ref{section:ionizing_background} for examples in which it does), 
but it does not appear to be essential.

%
%
\section{Summary}
\label{section:summary}

The 2012 Hubble Ultra Deep Field (UDF12) campaign, a
128-orbit {\it Hubble Space Telescope} ({\it HST}) program 
(GO 12498, PI: R. Ellis, as described in \citealt{ellis2012a}
and \citealt{koekemoer2012a}),
has acquired the deepest infrared WFC3/IR images ever 
taken with {\it HST}.  These observations have enabled
the first identification of galaxies at $8.5\le z \le 12$
in the Ultra Deep Field \citep{ellis2012a}, newly accurate
luminosity function determinations at redshifts $z\sim7-8$
\citep{schenker2012a,mclure2012a}, 
robust 
determinations of the ultraviolet spectral slopes of
galaxies at $z\sim7-8$ \citep{dunlop2012b}, and 
the first estimates of the 
luminosity function and spectral slopes at 
$z\sim9$ \citep{mclure2012a,dunlop2012b}.  Synthesizing these
constraints on high-redshift galaxy populations with the
recent 9-year {\it Wilkinson Microwave Anisotropy Probe}
({\it WMAP}) constraints on the electron scattering optical
depth generated by ionized cosmic hydrogen \citep{hinshaw2012a}
and previous
determinations of the galaxy luminosity function at redshifts
$4\lesssim z\lesssim6$ \citep{bouwens2007a}, we infer constraints
on the reionization history of the universe.  

First, we use the UV spectral slope $\beta = -2$ of high-redshift
galaxies measured by \citet{dunlop2012b} to constrain the
available \citet{bruzual2003a} stellar population models consistent
with the spectral character of galaxies at $z\sim7-9$. These
models motivate the adoption of a Lyman continuum photon production
rate per unit UV luminosity spectral density 
$\log \xiion = 25.2~\log~\xiionunit$; the data does not favor
a luminosity-dependent variation in the efficiency of Lyman
continuum photon production.  With this value of $\xiion$
for high-redshift galaxies,
and under reasonable assumptions for the Lyman continuum photon
escape fraction from galaxies and the clumping factor of 
intergalactic gas (as motivated by cosmological simulations),
we find that the currently observed galaxy population accessible
to the limiting depth of UDF12 ($\MUV<-17$ to $z\sim8$) cannot
simultaneously reionize the universe by $z\sim6$ and reproduce
the Thomson optical depth $\tau$ unless the abundance of star-forming
galaxies or the ionizing photon escape 
fraction {\it increases} beyond redshift $z\sim12$ from what is
currently observed.  

If we utilize constraints on the evolving
galaxy luminosity function at redshifts $4\lesssim z \lesssim9$
to extrapolate
down in luminosity, we find that the tension between the declining
abundance of star-forming galaxies and their stellar mass density
with redshift, the observed
requirement to reionize the universe by $z\sim6$, and reproducing
the large electron scattering optical depth $\tau\approx0.084$ 
is largely relieved if the galaxy population continues down to
$\MUV<-13$ and the epoch of galaxy formation continues to $z\sim12-15$.
Given the first identification of a $z\sim12$ candidate galaxy by
the UDF12 program, the prospect for high-redshift galaxies to 
reionize the universe is positive provided that the epoch of galaxy
formation extends to $z\gtrsim12$.  
Further 
observations by {\it HST} (e.g., the ``Frontier Fields'') and, ultimately, the 
{\it James Webb Space Telescope} will be required to answer these
questions more definitively.

\acknowledgments

We thank Gary Hinshaw and David Larson for pointing us
to the MCMC chains used in inferring the WMAP 9-year 
cosmological constraints.
BER is supported by Steward Observatory and the
University of Arizona College of Science.
SRF is partially supported by the David and Lucile Packard Foundation. 
US authors acknowledge financial support from the Space Telescope 
Science Institute under award HST-GO-12498.01-A. 
RJM acknowledges the support of the European Research Council via 
the award of a Consolidator Grant, and the support of the 
Leverhulme Trust via the award of a Philip Leverhulme research 
prize. JSD and RAAB acknowledge the support of the European 
Research Council via the award of an Advanced Grant to JSD. 
JSD also acknowledges the support of the Royal Society via a 
Wolfson Research Merit award. ABR and EFCL acknowledge the 
support of the UK Science \& Technology Facilities Council. 
SC acknowledges the support of the European Commission through 
the Marie Curie Initial Training Network ELIXIR. 
This work is based in part on observations made with the NASA/ESA 
Hubble Space Telescope, which is operated by the Association of 
Universities for Research in Astronomy, Inc, under NASA contract NAS5-26555. 

\bibliographystyle{apj}
\bibliography{ms}

\end{document}